\documentclass[aps,nofootinbib,noshowpacs,showkeys,preprintnumbers,amsmath,amssymb]{revtex4}
\def\bes{\begin{subequations}}
\def\ees{\end{subequations}}
\def\be{\begin{equation}}
\def\ee{\end{equation}}
\def\bea{\begin{eqnarray}}
\def\eea{\end{eqnarray}}
\def\ba{\begin{eqnarray}}
\def\ea{\end{eqnarray}}
\def\bear{\begin{array}}
\def\eear{\end{array}}
\def\p1sl{\displaystyle{\not}p_1}
\def\p2sl{\displaystyle{\not}p_2}

\newcommand{\K}{{\widetilde {\cal K}}}

\newcommand{\bG}{{\overline{\Gamma}}}

\newcommand{\Dst}{D^*}
\newcommand{\blam}{{\overline \lambda}}
\usepackage{mathtools}
\usepackage{graphics}
\usepackage{graphicx}
\usepackage{dcolumn}
\usepackage{bm}
\usepackage{epsfig}
\usepackage{graphicx}
\usepackage{multirow}
\usepackage{dcolumn}
\usepackage{graphicx,epsfig}%
\usepackage{color}

\begin{document}
\preprint{USM-TH-352}

\title{Sensitivity limits on heavy-light mixing $|U_{\mu N}|^2$ \\
from lepton number violating $B$ meson decays\footnote{v4: typos in Eqs.~(10), (11) and (17b) corrected; the correct expressions had been used in calculations, results unchanged.}}
\author{Gorazd Cveti\v{c}$^1$}
\email{gorazd.cvetic@usm.cl}
\author{C.~S.~Kim$^2$}
\email{cskim@yonsei.ac.kr}
\affiliation{$^1$
Department of Physics, Universidad T\'ecnica Federico Santa Mar\'ia, Valpara\'iso, Chile\\
$^3$Department of Physics and IPAP, Yonsei University, Seoul 120-749, Korea}

\date{\today}

\begin{abstract}
\noindent
We consider the lepton number violating decays $B \to  \mu^{\pm} \mu^{\pm} \pi^{\mp}$ and $B \to  D^{(*)} \mu^{\pm} \mu^{\pm} \pi^{\mp}$ which may be detected at LHCb and Belle-II experiments; and  $B \to \mu^{\pm} \mu^{\pm} e^{\mp} \nu$ and $B \to D^{(*)} \mu^{\pm} \mu^{\pm} e^{\mp} \nu$ decays which may be detected at Belle-II experiment. The projected total number of produced $B$ mesons is $4.8 \times 10^{12}$ at LHCb upgrade and $5 \times 10^{10}$ at Belle-II. For the case that the above decays are not detected, we deduce the new upper bounds (sensitivity limits) for the mixing parameter $|U_{\mu N}|^2$ of heavy sterile neutrino with sub-eV light neutrino, as a function of the sterile neutrino mass in the interval $1.75 \ {\rm GeV} < M_N < 5.0 \ {\rm GeV}$. We take into account the probability of decay of the sterile neutrino $N$ within the detector, taking as the effective detector length  $L=2.3 \ m$ at LCHb upgrade and $L=1 \ m$ at Belle-II.
In the interval $1.75 \ {\rm GeV} < M_N < 3 \ {\rm GeV}$, the most stringent bounds can be obtained with the decays $B \to \mu^{\pm} \mu^{\pm} \pi^{\mp}$ at LHCb upgrade. The sensitivity limits are expected to be in general more stringent at LHCb upgrade than at Belle-II, principally because the number of produced $B$ mesons in LHCb upgrade is expected to be by about two orders of magnitude larger than at Belle-II. We conclude that the LHCb upgrade and Belle-II experiments have the potential to either find a new heavy Majorana neutrino $N$, or to improve significantly the sensitivity limits (upper bounds) on the heavy-light mixing parameter $|U_{\mu N}|^2$, particularly in the mass range $1.75 \ {\rm GeV} < M_N < 3 \ {\rm GeV}$.
This work is a continuation and refinement of our previous work [Phys.\ Rev.\ D {\bf 94}, 053001 (2016); {\it ibid\/} {\bf 95}, 039901(E) (2017)] on the subject.
\end{abstract}
\pacs{14.60St, 13.20He}
\keywords{rare meson decays; sterile neutrino; mixing parameters of sterile neutrino and sub-eV neutrino }

\maketitle


\section{Introduction}
\label{intr}

The existence of sterile neutrinos has not been proven yet. However, their existence is suggested by various scenarios which can explain the detected differences of masses of the three known light neutrinos. Furthermore, most of such scenarios suggest that the neutrinos are Majorana fermions. Since Majorana fermions, unlike the Dirac fermions, are their own antiparticles, they can participate not just in the  lepton number conserving (LNC) processes, but also in the lepton number violating (LNV) processes. LNV processes are appreciable if the Majorana neutrinos are sufficiently massive. Various scenarios suggest that mixing of sterile neutrinos with the known Standard Model (SM) flavor neutrinos leads to neutrinos which are significantly heavier than the known light neutrinos. The main questions facing the neutrino physics beyond the SM are: (1) Are the neutrinos Majorana or Dirac? (2) How heavy are the new mass eigenstates  $N$? (3) What are the values of the  heavy-light mixing parameters $U_{\ell N}$, i.e., the mixing parameters of a massive $N$ neutrino with the SM flavor neutrinos $\nu_{\ell}$ ($\ell=e, \mu, \tau$)?

Whether the neutrinos are Majorana particles can be determined in neutrino experiments with various LNV processes. Among the most known such experiments are those with the neutrinoless double beta decay ($0\nu\beta\beta$) \cite{0NBB}, rare LNV decays of mesons \cite{RMDs,HKS,Atre,CDKK,CDK,CKZ,symm,Quint,Mand} and of $\tau$ lepton \cite{GKS,tau}, and specific scattering processes \cite{scatt1,scatt2,scatt3,KimLHC}.

Observation of neutrino oscillations \cite{Pontecorvo} can determine (small) mass differences between neutrinos, and thus prove that the neutrinos have mass. The neutrino oscillations of the SM flavor neutrinos have been observed \cite{oscatm,oscsol,oscnuc}. If sterile neutrinos exist and if their mixing with the SM flavor neutrinos leads to almost degenerate heavy neutrinos, also such neutrinos can oscillate among themselves \cite{Boya,CKZosc}.

The neutrino sector can also have CP violation \cite{oscCP}, which plays an important role in the leptogenesis \cite{Lepto}. Resonant CP violation of neutrinos appears when we have two heavy almost degenerate neutrinos. It can appear in scattering processes \cite{Pilaftsis}, in semileptonic rare meson decays \cite{CKZ2,DCK,symm}, and in purely leptonic rare meson decays \cite{CKZ,symm}. Among the models with almost degenerate heavy neutrinos are  the neutrino minimal standard model ($\nu$MSM) \cite{nuMSM,Shapo} and low-scale seesaw models \cite{lsseesaw}.

As mentioned, extended sectors of Majorana neutrinos appear in models which explain the very small masses of the three light neutrinos. Such models are the original seesaw models \cite{seesaw} (the heavy neutrinos there have masses $M_N \gg 1$ TeV), and seesaw models with heavy neutrinos with lower masses $M_N \sim 1$ TeV \cite{WWMMD}, and $M_N \sim 1$ GeV \cite{scatt2,nuMSM,HeAAS,KS,AMP,NSZ,Shrock}. In such models, the heavy-light mixing parameters are in general less suppressed than in the original seesaw models.

In this work, we will work in a generic framework where we have one massive neutrino $N$ which mixes with the  SM flavor neutrinos $\nu_{\ell}$ ($\ell=e, \mu, \tau$). We will evaluate the rates of some rare decays of $B$ mesons at the future \textcolor{black}{LHCb upgrade and}
Belle-II experiments, namely, the LNV decays with one on-shell Majorana massive neutrino $N$:  $B \to (D^{(*)}) \mu^{\pm} N \to  (D^{(*)}) \mu^{\pm} \mu^{\pm} X^{\mp}$, where $X^{\mp}$ is either a pion $\pi^{\mp}$, or a lepton-neutrino pair $\ell \nu_{\ell}$
\textcolor{black}{(this latter option only at Belle-II).}
This work is based on our previous work \cite{BdecBII}, but now the obtained results are more specific and directly applicable to the calculation of the sensitivity limits on the $|U_{\mu N}|^2$ mixing parameter, as a function of mass $M_N$, achievable
\textcolor{black}{at LHCb upgrade and at Belle-II,}
where the projected total number of produced $B$ mesons is
\textcolor{black}{$4.8 \times 10^{12}$ \cite{Sheldon} and $5 \times 10^{10}$ \cite{Belle-II}, respectively.}
Unlike in Ref.~\cite{BdecBII}, here we do not make any assumptions on the size of the probability $P_N$ of the produced neutrino $N$ to decay within the detector (in \cite{BdecBII} we assumed that either $P_N \approx 1$ or $P_N \ll 1$).
Detailed explanation on this issue is given in Sec.~\ref{sec:PN} and in Appendix \ref{appENpp}.

\textcolor{black}{Similar analyses for the upper bounds on $|U_{\mu N}|^2$ from the absence of the rare $B$-meson decays were made for the Belle-I mesurements in Ref.~\cite{BelleUB}, and for LHCb (run I) measurements in Refs.~\cite{LHCba1} and reconsideration thereof in Ref.~\cite{LHCba2}.}

In Sec.~\ref{sec:decw} we summarize the framework in which we work, and the decay widths which are relevant for the decay rates that we want to obtain. The summarized formulas  for these decay widths are presented in subsections of Sec. II and Appendix \ref{appNall}. In Sec.~\ref{sec:PN} we present the probability $P_N$ of the produced on-shell neutrino $N$ to decay within the detector, and the integration formulas which account for the effect of this probability on the effective rate for the mentioned LNV decays. In Appendix \ref{appENpp} we present detailed formulas for the Lorentz factors and the probabilities $P_N$ for the various considered decays.
In Sec.~\ref{sec:num} we present the results of the numerical evaluations, in the form of the obtained sensitivity limits on $|U_{\mu N}|^2$, as a function of $M_N$, that can be achieved by
\textcolor{black}{LHCb upgrade and}
Belle-II experiments. In Sec.~\ref{sec:concl} we discuss the obtained results and make conclusions.


\section{Decay widths for  $B \to (D^{(*)}) \ell_1 N \to  (D^{(*)}) \ell_1 \ell_2 X$}
\label{sec:decw}

Here we briefly summarize the results of Ref.~\cite{BdecBII} for the decay widths of the rare decays of $B$ mesons via on-shell sterile neutrino $N$. The on-shellness of $N$ implies the factorization
\be
 \Gamma \left( B \to (D^{(*)}) \ell_1 N \to (D^{(*)}) \ell_1 \ell_2 X \right)
  =  \Gamma \left( B \to (D^{(*)}) \ell_1 N \right) \frac{\Gamma(N \to \ell_2 X)}{\Gamma_N} \ .
\label{fact}
\ee
Here, $\ell_j$ ($j=1,2$) are generical names for charged leptons; later we will use $\ell_1 = \ell_2 = \mu^{\pm}$. The second factor on the right-hand side of Eq.~(\ref{fact}) represents the effect of the subsequent decay of the produced heavy on-shell neutrino $N$ into $\ell_2 + X$, where $X$ will be either a charged pion $\pi$, or a leptonic pair $\ell_3 \nu_3$.

The first factor in Eq.~(\ref{fact}), $\Gamma \left( B \to (D^{(*)}) \ell_1 N \right)$, is well known when no $D^{(*)}$ meson is produced; when $D^{(*)}$ is produced, this factor was obtained and evaluated in Ref.~\cite{BdecBII}. The formulas for this factor are summarized in subsections A-C, as well as some (here relevant) differential decay widths for these decays $B \to (D^{(*)}) \ell_1 N$. The second factor in Eq.~(\ref{fact}) includes the exclusive decay width $\Gamma(N \to \ell_2 X)$ which is well known, either for $X=\pi$ or $X=\ell_3 \nu_3$. For both cases, the expressions for these decay widths are summarized in subsections D-E. The denominator of the second factor in Eq.~(\ref{fact}), namely the total decay width $\Gamma_N$ of neutrino $N$, was evaluated numerically in \cite{CKZ2} for the case of Majorana $N$ (cf.~also \cite{symm} for the case of $N$ Majorana or Dirac); the expression for $\Gamma_N$ and its evaluation is presented in Appendix \ref{appNall}.

All the mentioned decay widths involve the (suppressed) heavy-light mixing parameters $U_{\ell N}$ ($\ell=e, \mu, \tau$) appearing in the coupling of the heavy $N$ neutrino with the $W$ boson and $\ell$ lepton. These parameters are part of the (extended) Pontecorvo-Maki-Nakagawa-Sakata (PMNS) matrix, i.e., the light flavor neutrino states $\nu_{\ell}$ (with flavor $\ell = e, \mu, \tau$) are the following combination of the three light mass eigenstates $\nu_k$ and of the heavy mass eigenstate $N$:
\be
\nu_{\ell} = \sum_{k=1}^3 U_{\ell \nu_k} \nu_k + U_{\ell N} N \ .
\label{mixN}
\ee


\subsection{Decay width $\Gamma(B \to \ell_1 N)$}
\label{subs:GBellN}

The decay width for the process  $B \to \ell_1 N$, where $\ell_1$ is a charged lepton ($\ell_1=e, \mu, \tau$) and $N$ is a (massive) neutrino, is
\be
\Gamma(B^{\pm} \to \ell_1^{\pm} N) =  |U_{\ell_1 N}|^2
\bG (B^{\pm} \to \ell_1^{\pm} N) \  ,
\label{GBlN}
\ee
where the canonical decay width $\bG$, i.e., the part without the heavy-light mixing factor, is
\be
\bG (B^{\pm} \to \ell_1^{\pm} N) =
\frac{G_F^2 f_{B}^2}{8 \pi} |V_{u b}|^2 M_{B}^3 \lambda^{1/2}(1,y_N,y_1)
\left[ (1 - y_N) y_N + y_1 (1 + 2 y_N - y_1) \right] \ .
\label{bGBlN}
\ee
Here, $G_F$ is the Fermi coupling constant ($G_F = 1.1664 \times 10^{-5} \ {\rm GeV}^{-2}$), $f_{B}$ is the decay constant of the $B$-meson,
$V_{u b}$ its CKM matrix element, and in the mass dependent parts the following notations are used:
\bes
\label{notylam}
\bea
y_N &=& \frac{M_N^2}{M_B^2} \ , \qquad y_1 = \frac{M_1^2}{M_B^2} \ ,
\label{yNyell}
\\
\lambda^{1/2}(x,y,z) &=& \left[ x^2 + y^2 + z^2 - 2 x y - 2 y z - 2 z x \right]^{1/2}.
\label{lam}
\eea
\ees
We denote the mass of $\ell_1$ as $M_1$ throughout this paper. We use the values $|V_{ub}|=0.00409$ and $f_B=0.1871$ GeV \cite{PDG2016} (cf.~also \cite{Kangetal}).

\subsection{Decay width $\Gamma(B \to D \ell_1 N)$}
\label{subs:GBDellN}

\begin{figure}[htb]
\centering\includegraphics[width=90mm]{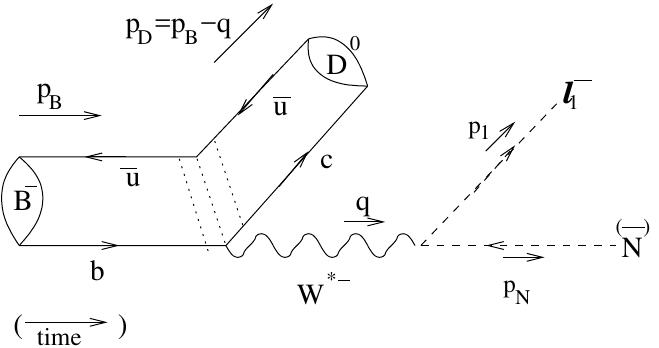}
\caption{Schematical representation of the decay $B^- \to D^0 \ell_1^- {\bar N}$.}
\label{FigBDW}
\end{figure}

We now consider the decay $B \to D \ell_1 N$, cf.~Fig.~\ref{FigBDW}.
For the general case of a massive neutrino $N$ (and a massive charged lepton $\ell_1$), the general expression for the decay width of the process $B \to D \ell_1 N$ was obtained in Ref.~\cite{BdecBII}. There, the differential decay width $d \Gamma(B^- \to D^0 \ell_1^- {\bar N})/d q^2$ was presented.

Here we present the ``more differential'' cross section $d \Gamma(B^- \to D^0 \ell_1^- {\bar N})/(d q^2 d \Omega_{{\hat q}'} d \Omega_{{\hat p}_1})$, which is needed for calculation of the effective (true) branching ratio $ {\rm Br}_{\rm eff}(B \to D \ell_1 N \to D \ell_1 \ell_2 X)$ of Eq.~(\ref{Breff}).
The differential of the decay width is
\be
d \Gamma(B^- \to D^0 \ell^-_1 N) = \frac{1}{2 M_B} \frac{1}{(2 \pi)^5} d_3 |{\cal T}|^2 \ ,
\label{BDNl}
\ee
where $d_3$ is the differential for the three-particle final phase space
\bea
d_3 & = & \frac{d^3 {\vec p}_D}{2 E_{D}({\vec p}_D)}
 \frac{d^3 {\vec p}_1}{2 E_{\ell_1}({\vec p}_1)}
 \frac{d^3 {\vec p}_N}{2 E_N({\vec p}_N)}
 \delta^{(4)} \left( p_B - p_D - p_1 - p_N \right)
 \nonumber\\
 & = & d_2 \left( B^- \to D^0(p_D) W^*(q) \right) d q^2 d_2 \left( W^*(q) \to \ell_1(p_1) {\overline N}(p_N) \right) \ ,
\label{d3}
\eea
and the two-particle final phase space differentials are
\bes
\label{d2}
\bea
d_2(B^- \to D^0(p_D) W^*(q)) & = & \frac{1}{8} \lambda^{1/2} \left( 1, \frac{M_D^2}{M_B^2}, \frac{q^2}{M_B^2} \right) d \Omega_{{\hat q}'},
\label{d2BDW}
\\
d_2(W^*(q) \to \ell^-_1(p_1) {\overline N}(p_N)) & = & \frac{1}{8} \lambda^{1/2} \left( 1, \frac{M_1^2}{q^2}, \frac{M_N^2}{q^2} \right) d \Omega_{{\hat p}_1}.
\label{d2WellN}
\eea
\ees
The decay amplitude ${\cal T}$ appearing in Eq.~(\ref{BDNl}) is
\be
{\cal T}  =  U_{\ell_1 N} V_{c b} \frac{G_F}{\sqrt{2}} \left[{\overline u}_{(\ell_1)}(p_1) \gamma_{\mu} (1 - \gamma_5) v_{(N)}(p_N) \right]
\left\{ \left[ (2 p_D + q)^{\mu} - \frac{(M_B^2-M_D^2)}{q^2} q^{\mu} \right] F_1(q^2) + \frac{(M_B^2-M_D^2)}{q^2} q^{\mu} F_0(q^2) \right\},
\label{TBDNl}
\ee
where $F_1(q^2)$ and $F_0(q^2)$ are the form factors of the
$B$-$D$ transition, and we consider them to be real.

In terms of the reduced canonical decay amplitude ${\widetilde {\cal T}}$ defined via the relation
\be
| {\cal T} |^2 = 4 |U_{\ell_1 N}|^2 |V_{c b}|^2 G_F^2 |{\widetilde {\cal T}}|^2,
\label{tildeT}
\ee
we can then express the differential decay width (\ref{BDNl}) in a somewhat more explicit form,\footnote{In v3, the right-hand side of Eq.~(\ref{dGBDlN}) has a typo (a superfluous factor $1/4$), but in the calculations the correct expression was used.}
\bea
\frac{d \Gamma(B^- \to D^0 \ell^-_1 N)}{d q^2 d \Omega_{{\hat q}'} d \Omega_{{\hat p}_1}} & = & \frac{|U_{\ell_1 N}|^2 |V_{c b}|^2 G_F^2}{M_B (4 \pi)^5} |{\widetilde {\cal T}}|^2  \lambda^{1/2} \left( 1, \frac{M_D^2}{M_B^2}, \frac{q^2}{M_B^2} \right)  \lambda^{1/2} \left( 1, \frac{M_1^2}{q^2}, \frac{M_N^2}{q^2} \right),
\label{dGBDlN}
\eea
where ${\hat p}_1$ is the direction of  $\ell^-_1$ in the $W^*$-rest frame ($\Sigma$), and ${\hat q}'$ is the direction of $W^{*-}$ ($\ell^-_1 N$ pair) in the $B$-rest frame ($\Sigma'$). We use the expression (\ref{TBDNl}) for the decay amplitude, and calculate the square of its absolute magnitude, $|{\cal T}|^2$, summing over the helicities of the final particles. We then obtain for the square of the reduced canonical amplitude, $|{\widetilde {\cal T}}|^2$, introduced via Eq.~(\ref{tildeT}), the following expression:
\bea
|{\widetilde {\cal T}}|^2 &=&
\frac{1}{q^2} F_1(q^2) (F_0(q^2)-F_1(q^2)) \left(M_B^2-M_D^2\right)
   {\bigg [}  M_1^2 \left(-4 (\cos \theta_1 |{\vec p}_D| |{\vec p}_N|+p_D^0
     p_1^0)+2 M_B^2-2 M_D^2+2 M_N^2-q^2\right)
\nonumber\\
&&
     +M_N^2 \left(4 (\cos \theta_1 |{\vec p}_D|
   |{\vec p}_N|+p_D^0
   p_1^0)-M_N^2+q^2\right)-M_1^4 {\bigg ]}
\nonumber\\
&&
-\frac{1}{2} F_1(q^2)^2 {\bigg [}M_1^2
   \left( 8 (\cos \theta_1 |{\vec p}_D|
   |{\vec p}_N|+p_D^0 p_1^0)-4 M_B^2-2 M_N^2+3
   q^2\right)
   -8 M_B^2 (\cos \theta_1 |{\vec p}_D| |{\vec p}_N|+
   p_D^0 p_1^0)
 \nonumber\\
&&
+M_D^2 \left(8 (\cos \theta_1 |{\vec p}_D|
|{\vec p}_N|+p_D^0 p_1^0)-4 M_N^2+4 q^2\right)
-8 M_N^2 (\cos \theta_1 |{\vec p}_D| |{\vec p}_N|+p_D^0 p_1^0)
+8q^2 (\cos \theta_1 |{\vec p}_D| |{\vec p}_N|+p_D^0 p_1^0)
 \nonumber\\
&&
+16 (\cos \theta_1 |{\vec p}_D| |{\vec p}_N|+p_D^0
   p_1^0)^2+M_1^4+M_N^4-M_N^2
   q^2 {\bigg ]}
\nonumber\\
&&
+\frac{1}{2 (q^2)^2}
(F_0(q^2)-F_1(q^2))^2 \left(M_B^2-M_D^2\right)^2 \left[-M_1^4+M_1^2 \left(2
     M_N^2+q^2\right)-M_N^4+M_N^2 q^2\right] \ .
\label{tildeTexp}
\eea
Here, we denoted as $p_1$ the 4-momentum of $\ell_1$ (in $W^{*}$-rest frame $\Sigma$), and $\theta_1$ is the angle between ${\vec p}_1$ and ${\hat z} = {\hat q}'$. We also used in Eq.~(\ref{tildeTexp})  the following quantities:
\bes
\label{vecpo0}
\bea
|{\vec p}_N| = |{\vec p}_1| & = & \frac{1}{2} \sqrt{q^2}  \; \lambda^{1/2} \left( 1, \frac{M_1^2}{q^2}, \frac{M_N^2}{q^2} \right),
\label{vecpN}
\\
|{\vec p}_D| & = & \frac{M_B^2}{2 \sqrt{q^2}} \; \lambda^{1/2} \left( 1, \frac{M_D^2}{M_B^2}, \frac{q^2}{M_B^2} \right) = \frac{M_B |{\vec {q'}}|}{\sqrt{q^2}},
\label{vecpD}
\\
p_1^0 & = & \frac{1}{2 \sqrt{q^2}} (q^2 - M_N^2 + M_1^2),
\label{p10}
\\
p_D^0 & = & \frac{1}{2 \sqrt{q^2}} (M_B^2 - M_D^2 - q^2).
\label{pD0}
\eea
\ees
They are all in the $W^*$-rest frame ($\Sigma$). We can see from these expressions that the absolute square of the reduced canonical amplitude, $|{\widetilde {\cal T}}|^2$, and thus the differential decay width (\ref{dGBDlN}), depend only on the variables $q^2$ (square of the invariant mass of $W^{*}$) and on $\cos \theta_1$ [note: $d \Omega_{{\hat p}_1} = d \phi_1 d (\cos \theta_1$)]. They are thus independent of the direction ${\hat q}'$, i.e., of the direction of $W^*$ in the $B$-rest frame.

The expressions (\ref{tildeTexp}) and (\ref{dGBDlN}) contain two form factors, $F_1$ and $F_0$. The form factor $F_1(q^2)$ is well known \cite{CLN} and can be expressed in terms of a variable $w(q^2)$
\bes
\label{wz}
\bea
w & = & \frac{(M_B^2 + M_D^2 - q^2)}{2 M_B M_D} \ ,
\label{w}
\\
z(w) & = & \frac{\sqrt{w+1} - \sqrt{2}}{\sqrt{w+1} + \sqrt{2}} \ .
\label{z}
\eea
\ees
According to Ref.~\cite{CLN}, $F_1(q^2)$ has the following power expansion in $z(w(q^2))$:
\be
F_1(q^2) = F_1(w=1) \left( 1 - 8 \rho^2 z(w) + (51 \rho^2 - 10) z(w)^2 - (252 \rho^2 - 84) z(w)^3 \right) \ .
\label{CLNF1}
\ee
The free parameters $\rho^2$ and $F_1(w=1)$ in this expansion have been determined by the Belle Collaboration, Ref.~\cite{Belle1}
\bes
\label{rho2F1max}
\bea
\rho^2 &= & 1.09 \pm 0.05 \ ,
\label{rho2}
\\
|V_{cb}| F_1(w=1) &=& (48.14 \pm 1.56) \times 10^{-3} \ .
\label{F1max}
\eea
\ees
In our numerical evaluations we use the above central values, and $|V_{cb}|=40.12 \times 10^{-3}$ \cite{Belle1}.

The form factor $F_0(q^2)$ is not well known at present, principally because it contributes only when the masses of $N$ and $\ell_1$ are not very small as can be deduced from Eq.~(\ref{tildeTexp}).\footnote{It can be checked that the difference $[ |{\widetilde {\cal T}}|^2 - |{\widetilde {\cal T}}|^2(F_0 \mapsto 0)]$ is zero when $M_1=M_N=0$.} In our case $F_0(q^2)$ is important, and it was presented in Ref.~\cite{BdecBII} by using  the truncated expansion for $F_0$ in powers of $w(q^2) - 1$ of Ref.~\cite{CaNeu}
\bes
\label{F0}
\bea
F_0(q^2) & = & \frac{(M_B+M_D)}{2 \sqrt{M_B M_D}}
\left[ 1 - \frac{q^2}{(M_B+M_D)^2} \right] f_0(w(q^2)) \ ,
\label{F0a}
\\
f_0(w) & \approx & f_0(w=1) \left[ 1 - {\rho}_0^2 (w - 1) + (0.72 \rho_0^2 - 0.09) (w - 1)^2 \right] \ .
\label{F0b}
\eea
\ees
Here,\footnote{In v3, as well as in our previous work \cite{BdecBII}, the expression (\ref{F0b}) was written with a typo [$+{\rho}_0^2 (w - 1)$ instead of $- {\rho}_0^2 (w - 1)$], but the correct expression was used in the calculations here and in \cite{BdecBII}.}
 we use the value $f_0(w=1) \approx 1.02$ \cite{NeuPRps,CaNeu} which is obtained from the heavy quark limit. The other free parameter $\rho_0$ in Eq.~(\ref{F0b}) is then fixed by requiring the absence of spurious poles at $q^2=0$: $F_0(0)=F_1(0)$ ($\approx 0.690$). This yields the value $\rho_0^2 \approx 1.102$ and $(0.72 \rho_0^2 - 0.09) \approx 0.704$.

For the curves of these form factors $F_1(q^2)$ and $F_0(Q^2)$, as a function of positive $q^2$, we refer to Ref.~\cite{BdecBII} (Fig.~2 there).

\subsection{Decay width $\Gamma(B \to D^{*} \ell_1 N)$}
\label{subs:GBDstellN}

We now consider the decay $B \to D^{*} \ell_1 N$, i.e., the same type of decay as in the previous Sec.~\ref{subs:GBDellN}, but now instead of the (pseudoscalar) $D$ meson we have vector meson $D^{*}$. The expressions for the (differential) decay widths are now more complicated, because $D^{*}$ is a vector particle. For the case of massive neutrino $N$ (and massive lepton $\ell_1$), these expressions were obtained in Ref.~\cite{BdecBII}, using the approach of Ref.~\cite{GiSi}. The needed differential decay width, after summation over helicities and over the three polarizations of $D^{*}$, turns out to be \cite{BdecBII}
 \bea
  \frac{d \Gamma}{dq^2 d \Omega_{{\hat q}'} d \Omega_{{\hat p}_1}} & = &
  \frac{1}{8^4 \pi^5} \frac{|U_{\ell_1 N}|^2 |V_{cb}|^2 G_F^2}{M_B^2}
  \blam^{1/2} 2 |{\vec {q'}}|  q^2 {\bigg \{}
\left[2 \left(1 - \frac{(M_N^2+M_1^2)}{q^2} \right) -
  \blam \sin^2 \theta_1  \right]
\left( ({\bar H_{+1}})^2 + ({\bar H_{-1}})^2 \right)
\nonumber\\
&& - \eta \; 2 \blam^{1/2} \cos \theta_1
\left( ({\bar H_{+1}})^2 - ({\bar H_{-1}})^2 \right)
+ 2 \left[ \left(1 - \frac{(M_N^2+M_1^2)}{q^2} \right) - \blam \cos^2 \theta_1 \right] ({\bar H^3})^2
\nonumber\\
&& +
4 \left( \frac{M_N^2-M_1^2}{q^2} \right) \blam^{1/2} \cos \theta_1 {\bar H^0}{\bar H^3}
+ 2 \left[ - \left(\frac{M_N^2-M_1^2}{q^2} \right)^2 + \frac{(M_N^2+M_1^2)}{q^2}
  \right] ({\bar H^0})^2 {\bigg \}} \ .
    \label{dGdq2domdom2}
\eea
Here, the factor $\eta=\pm 1$ appears at one term proportional to $\cos \theta_1$; $\eta=+1$ if $\ell^-_1$ is produced, and $\eta=-1$ if $\ell^+_1$ is produced.\footnote{The quantity (\ref{dGdq2domdom2}) is written in Ref.~\cite{BdecBII} in Eq.~(C19) for the case $\eta=-1$; the quantity $d \Gamma/d q^2$ used there is independent of $\eta$.} Further, the following notations are used:
\bes
\label{notBDstellN}
\bea
|{\vec {q'}}| &=& \frac{1}{2} M_B \lambda^{1/2} \left( 1, \frac{ M_{\Dst}^2}{M_B^2}, \frac{q^2}{M_B^2} \right),
\label{magq}
\\
\blam &\equiv& \lambda \left( 1, \frac{M_1^2}{q^2}, \frac{M_N^2}{q^2} \right) \ ,
\label{blam}
\eea
\ees
and ${\bar H}_{\pm 1}$, ${\bar H^0}$ and ${\bar H^3}$ are expressions containing the form factors $V$ and $A_j$ ($j=0,1,2,3$) appearing in the $B$-$D^{*}$ matrix elements
\bes
 \label{bHs}
 \bea
 {\bar H_{\pm 1}} &=& (M_B+M_{\Dst}) A_1(q^2) \mp V(q^2) \frac{|{\vec {q'}}| 2 M_B}{(M_B+M_{\Dst})} \ ,
 \label{bHpm}
 \\
 {\bar H^3} & = & \frac{M_B^2}{2 M_{\Dst} \sqrt{q^2}} \left[
       (M_B+M_{\Dst}) A_1(q^2) \left(1 - \frac{(q^2+M_{\Dst}^2)}{M_B^2} \right)
         - 4 A_2(q^2) \frac{|{\vec {q'}}|^2}{(M_B+M_{\Dst})} \right] \ ,
 \label{bH3}
 \\
 {\bar H^0} & = & \frac{M_B |{\vec {q'}}|}{M_{\Dst} \sqrt{q^2}} \left[
        (M_B+M_{\Dst}) A_1(q^2)  - (M_B- M_{\Dst}) A_2(q^2) + 2 M_{\Dst} \left( A_0(q^2) - A_3(q^2) \right) \right] \ .
 \label{bH0}
 \eea
\ees
$A_3$ form factor is not independent, it is a linear combination of $A_1$ and $A_2$
\be
A_3(q^2) = \frac{(M_B+M_{\Dst})}{2 M_{\Dst}} A_1(q^2) -
\frac{(M_B-M_{\Dst})}{2 M_{\Dst}} A_2(q^2) \ .
\label{A3}
\ee
Among the other four form factors, three ($V$, $A_1$ and $A_2$) are well known, they were recently determined to a high precision  \cite{Belle2} in terms of the parametrization of Ref.~\cite{CLN}
\bes
\label{A1VA2}
\bea
A_1(q^2) & = & \frac{1}{2} R_* (w+1) F_*(1) \left[ 1 - 8 \rho_*^2 z(w) + (53 \rho_*^2 - 15) z(w)^2 - (231 \rho_*^2 - 91) z(w)^3 \right] \ ,
\label{A1}
\\
V(q^2) & = & A_1(q^2)  \frac{2}{R_*^2 (w+1)} \left[ R_1(1) - 0.12 (w-1) + 0.05 (w-1)^2 \right] \ ,
\label{V}
\\
A_2(q^2) & = & A_1(q^2) \frac{2}{R_*^2 (w+1)} \left[ R_2(1) + 0.11 (w-1) - 0.06 (w-1)^2 \right] \ .
\label{A2}
\eea
\ees
The notation  $R_* = 2 \sqrt{ M_B M_{\Dst}}/(M_B+M_{\Dst})$ is used here, and $w=w(q^2)$ and $z=z(w(q^2))$ are given in Eqs.~(\ref{wz}) (with $M_D \mapsto M_{D^{*}}$). The values of the three parameters in Eqs.~(\ref{A1VA2}) were determined in Ref.~\cite{Belle2}
\bes
\label{paramsDst}
\bea
\rho_*^2 & = & 1.214(\pm 0.035) \ , \qquad 10^3 F_*(1) |V_{cb}| = 34.6(\pm 1.0) \ ,
\label{rhostFst}
\\
R_1(1) & = &1.401(\pm 0.038) \ , \qquad R_2(1) = 0.864(\pm 0.025) \ .
\label{R1R2}
\eea
\ees
We use the central values in the present work.

The form factor $A_0$, on the other hand, is not well known. It is relevant only if the masses of $N$ or $\ell_1$ are nonnegligible, which is the case here. Employing the heavy quark limit relations between $A_1$ and $A_2$, the relation (\ref{A3}) gives a relation between $A_2$ and $A_3$. Using this relation in the heavy quark limit relation $A_0 \approx A_2$, we then obtain the following approximation for the form factor $A_0$ in terms of $A_3$:
\be
A_0(q^2) \approx A_3(q^2)/\left[1 - \frac{q^2}{2 M_{\Dst} (M_B+M_{\Dst})} \right]
= \frac{(M_B+M_{\Dst})^2}{\left( 2 M_{\Dst} (M_B+M_{\Dst}) - q^2 \right)}
\left( 1 - \frac{(M_B-M_{\Dst})}{(M_B+M_{\Dst})} \frac{A_2(q^2)}{A_1(q^2)} \right) A_1(q^2)
\ ,
\label{A0appr}
\ee
This relation satisfies the relation $A_0(0)=A_3(0)$ which is obligatory since it reflects the absence of the pole at $q^2=0$ in the $B$-$D^{*}$ matrix elements. We refer for any further details on these points to Ref.~\cite{BdecBII}.



\subsection{Decay width for $N \to \ell^{\pm} \pi^{\mp}$}
  \label{subs:Nellpi}

The decay width $\Gamma(N \to \ell^{\pm} \pi^{\mp})$ is proportional to the heavy-light mixing factor $|U_{\ell N}|^2$
\be
\Gamma(N \to \ell^{\pm} \pi^{\mp}) = |U_{\ell N}|^2 \bG(N \to \ell^{\pm} \pi^{\mp}) \ .
\label{GNlPi}
\ee
Here, the canonical decay width $\bG$ is (e.g., cf.~Refs.~\cite{CDKK,CKZ2,symm,CKZosc})
\be
\bG(N \to \ell^{\pm} \pi^{\mp}) =
\frac{1}{16 \pi} |V_{u d}|^2 G_F^2 f_{\pi}^2 M_N^3 \lambda^{1/2}(1, x_{\pi}, x_{\ell})
\left[ 1 - x_{\pi} - 2 x_{\ell} - x_{\ell}  (x_{\pi}-x_{\ell}) \right] \ ,
\label{bGNlPi}
\ee
where $f_{\pi}$ ($\approx 0.1304$ GeV) is the decay constant of pion,
and we use the notations
\be
x_{\pi} = \frac{M_{\pi}^2}{M_N^2} \ , \qquad x_{\ell}=\frac{M_{\ell}^2}{M_N^2} .
\label{xPixell}
\ee

\subsection{Decay width for $N \to \ell_2 \ell_3 \nu$}
\label{subs:Nellellnu}

If the heavy neutrino $N$ is produced by the decay $B \to (D^{(*)}) \ell_1^{\pm} N$, the neutrino can decay into various leptonic channels $\ell_2 \ell_3 \nu$.
We can have the leptonic decays of $N$ of the lepton number conserving (LNC) type  $N \to \ell_2^{\mp} \ell_3^{\pm} \nu_{\ell_3}$, and of the lepton number violating (LNV) type $N \to \ell_3^{\pm} \ell_2^{\mp} \nu_{\ell_2}$
\bes
\label{GNllnu}
\bea
\Gamma^{\rm (LNC)}(N \to \ell_2^{\mp} \ell_3^{\pm} \nu_{\ell_3}) & = & |U_{\ell_2 N}|^2 \bG (N \to \ell_2 \ell_3 \nu) \ ,
\label{GNllnu.LC}
\\
 \Gamma^{\rm (LNV)}(N \to \ell_3^{\pm} \ell_2^{\mp} \nu_{\ell_2}) & = & |U_{\ell_3 N}|^2 \bG (N \to \ell_2 \ell_3 \nu) \ .
\label{GNllnu.LV}
\eea
\ees
Here, the charged leptons can be $\mu, e$ or $\tau$. The canonical decay widths $\bG(N \to \ell_2 \ell_3 \nu)$ have in the general case (with masses of leptons) the following form \cite{CDKK,CKZ,symm}:
\be
 {\overline \Gamma}(N \to \ell_2 \ell_3 \nu) =  \frac{G_F^2 M_N^2}{192 \pi^3}
{\cal F}(x_2,x_3) \ ,
\label{bGNllnu}
\ee
where we denoted $x_j = M_j^2/M_N^2$ ($M_j$ is the mass of $\ell_j$), and the function ${\cal F}$ is \cite{CKZ}
\bea
\lefteqn{
{\cal F}(x_2,x_3) =
{\Bigg \{}
\lambda^{1/2} (1, x_2, x_3) {\Big [} (1 + x_2) (1 -8 x_2 + x_2^2)  -
x_3 (7 - 12 x_2 + 7 x_2^2)
}
\nonumber\\
&&
- 7 x_3^2 (1 + x_2)  + x_3^3 {\Big ]}
- 24 (1 - x_3^2) x_2^2 \ln 2
\nonumber\\
&&
+  12 {\bigg [} - x_2^2 (1 - x_3^2) \ln x_2
+ (2 x_2^2 -x_3^2 (1 + x_2^2)) \ln (1 + x_2
+ \lambda^{1/2} (1, x_2, x_3)  - x_3)
\nonumber\\
&&
+ x_3^2 (1 - x_2^2)
\ln \left( \frac{(1 - x_2)^2 + (1-x_2) \lambda^{1/2} (1, x_2, x_3) - x_3 (1+x_2)}{x_3}
\right) {\bigg ]}
{\Bigg \}} .
\label{calF}
\eea
The function ${\cal F}$ is symmetric under the exchange of the two arguments. When one lepton is massless (or almost massless, i.e., lepton $e$), this expression reduces to the well-known result
\be
{\cal F}(x,0) =  {\cal F}(0,x)= f(x) = 1 - 8 x + 8 x^3 - x^4 - 12 x^2 \ln x \ .
\label{fx}
\ee

\section{Decay probability of heavy  neutrino in the detector; effective branching ratio}
\label{sec:PN}

If all the neutrinos $N$ decay within the detector with probability one, then the decay width Eq.~(\ref{fact}) is also the effective (true) decay width, and the effective branching ratio is obtained by dividing it by the decay width of the $B$ meson $\Gamma_B$. However, since the neutrino $N$ is weakly coupled to SM particles, it often does not decay within the detector and, consequently, the mentioned decays $B \to (D^{(*)}) \ell_1 \ell_2 X$ are not observed although $N$ may be produced in the $B$-decays. The effect of the decay of $N$ can be accounted for by multiplying the above decay width Eq.~(\ref{fact}) by the decay (nonsurvival) probability $P_N$ of $N$ within the detector
\be
P_N = 1 - \exp \left[ - \frac{L}{\tau_N \gamma_N \beta_N} \right]
= 1 - \exp \left[ - \frac{L \Gamma_N}{\gamma_N \beta_N} \right]
\label{PN}
\ee
where $L$ is the maximum possible flight length of $N$ within the detector, $\beta_N$ is the velocity of $N$ in the lab frame, $\tau_N = 1/\Gamma_N$ is the lifetime of $N$ in its rest frame, and $\gamma_N =(1 - \beta_N^2)^{-1/2}$ is the Lorentz time dilation factor \cite{CDK,CKZ,CKZ2,symm,scatt3,CERN-SPS,commKim,Gronau}.

For Belle-II, the $B$ meson pairs will be produced in SuperKEKB in central collisions of $e^-(p_1)$ and $e^+(p_2)$, which will produce a moving $\Upsilon(4S)$, the latter decaying into a $B$ meson pair (either $B^+ B^-$ or $B^0 {\bar B}^0$). In the lab frame, the $e^{\pm}$ have the momenta
\be
p_j = \left( E_j,0,0,(-1)^{j+1} E_j \right) \qquad (j=1,2),
\label{E1E2}
\ee
with the values $E_1=7.007$ GeV and $E_2 = 3.993$ GeV. This then produces the invariant mass $(p_1+p_2)^2 = M^2_{\Upsilon(4 S)}$, where $M_{\Upsilon(4 S)}=10.579$ GeV \cite{PDG2016}. The kinetic energy of the produced $\Upsilon(4 S)$ is $K_{\Upsilon}=E_1 + E_2 -M_{\Upsilon(4 S)} = 0.421$ GeV, which is semirelativistic, leading to the Lorentz factor in the lab frame
\be
\gamma_{\Upsilon}  =  \frac{(E_1+E_2)}{M_{\Upsilon(4 S)}} = 1.0398
\; \Rightarrow \;  \beta_{\Upsilon}  =  (1 - 1/\gamma_{\Upsilon}^2)^{1/2} = 0.274 \ .
\label{gammabetaU}
\ee
When $\Upsilon(4 S)$ produces $B$ meson pair, the kinetic energy of the produced $B$ mesons is about $0.010$ GeV in the $\Upsilon(4 S)$-rest frame, which is negligible. Therefore, we consider the velocity of the produced $B$ mesons in the lab frame to be the same as the velocity of  $\Upsilon(4 S)$
\be
\beta_B = \beta_{\Upsilon} = 0.274, \qquad
\gamma_B = \gamma_{\Upsilon} = 1.0398, \qquad
(p_B)_{\rm lab}=M_B \beta_B \gamma_B = 1.504 \ {\rm GeV}.
\label{gammabetaB}
\ee
In the decays $B \to D^{(*)} \ell_1 N$, we will denote the rest frame of the off-shell $W^*$ (i.e., of $\ell_1 N$ pair) as $\Sigma$; the $B$-rest frame as $\Sigma'$; the laboratory frame as $\Sigma''$. With these notations, the effective (true) branching ratio is calculated
\bea
{\rm Br}_{\rm eff}(B \to D^{(*)} \ell_1 N \to D^{(*)} \ell_1 \ell_2 X) & = &
    \int d q^2 \int d \Omega_{{\hat q}'} \int d \Omega_{{\hat p}_1}
    \frac{d \Gamma(B \to D^{(*)} \ell_1 N)}{ d q^2 d \Omega_{{\hat q}'}  d \Omega_{{\hat p}_1}} \frac{ \Gamma(N \to \ell_2 X) }{\Gamma_N \Gamma_B}
\nonumber\\
&& \times \left\{ 1 - \exp \left[- \frac{L \Gamma_N}{\sqrt{ \left(E''_N(q^2;{\hat q}',{\hat p}_{1})/M_N \right)^2 - 1 }} \right] \right\},
\label{Breff}
\eea
where in the denominator inside the exponent we have the Lorentz factor
\be
\beta_N^{''} \gamma_N^{''} = \sqrt{ \left(E''_N(q^2;{\hat q}',{\hat p}_{1})/M_N \right)^2 - 1 } \ ,
\label{bNgNpp}
\ee
in the laboratory frame, which is a function of $W^*$ ($=\ell_1 N$) momentum $q'$ (in the $B$-rest frame)\footnote{Note that ${q'}^2=q^2$ is frame independent.}
and of the direction ${\hat p}_{1}$ of the momentum $p_{1}$ of the produced charged lepton $\ell_1$ (in the $W^*$-rest frame). The expression (\ref{bNgNpp}) as an explicit function of $q^2$, ${\hat q}^{'}$ and ${\hat p}_1$ is derived in Appendix \ref{appENpp}. It depends on the angle $\theta_q$ between the direction of ${\hat \beta}_B$ (in the lab frame $\Sigma''$) and ${\hat q}'$ of $W^*$ (in the $B$-rest frame $\Sigma'$), as well as on the spherical angles $\theta_1$ and $\phi_1$ of the vector ${\vec p}_1$ of $\ell_1$ in the $W^*$-rest ($\Sigma$) frame, in a specific 3-dimensional system of coordinates in the frame $\Sigma$ (cf.~Fig.~\ref{Figthqth1} in Appendix \ref{appENpp}). On the other hand, the differential decay width $d \Gamma(B \to D^{(*)} \ell_1 N)/( d q^2 d \Omega_{{\hat q}'} d \Omega_{{\hat p}_1})$ depends only on $q^2$ and $\theta_1$, as shown in subsections \ref{subs:GBDellN}-\ref{subs:GBDstellN}. Due to the mentioned dependence in the decay (nonsurvival) factor $P_N$, integration over these momenta is needed, as indicated in Eq.~(\ref{Breff}). The differential decay widths $d \Gamma(B \to D^{(*)} \ell_1 N)/(d q^2 d \Omega_{{\hat q}'} d \Omega_{{\hat p}_1})$ are given in subsections \ref{subs:GBDellN}-\ref{subs:GBDstellN}. All this implies that the integration Eq.~(\ref{Breff}) has the following form:
\be
\int_{(M_N+M_1)^2}^{(M_B - M_{D^{(*)}})^2} d q^2 2 \pi \int_{-1}^{+1}  d (\cos \theta_q) \int_{-1}^{+1} d (\cos \theta_1) \int_0^{2 \pi} d \phi_1 f(q^2, \theta_q, \theta_1, \phi_1) .
\label{integrbounds}
\ee

If no mesons $D^{(*)}$ are produced in the decays, then the differential decay width is even simpler, as it depends only on the direction ${\hat {p}}'_N$ of the on-shell $N$ in the $B$-rest frame, and the expression (\ref{Breff}) simplifies
\bea
{\rm Br}_{\rm eff}(B \to \ell_1 N \to \ell_1 \ell_2 X) & = &
    \int d \Omega_{{\hat p}'_N}
    \frac{d \Gamma(B \to \ell_1 N)}{ d \Omega_{{\hat p}'_N} } \frac{ \Gamma(N \to \ell_2 X) }{\Gamma_N \Gamma_B}
\nonumber\\
&&\times \left\{ 1 - \exp \left[- \frac{L \Gamma_N}{\sqrt{ \left(E''_N({\hat p}'_N)/M_N \right)^2 - 1 }} \right] \right\}.
\label{BreffnoD}
\eea
The differential decay width is $d \Gamma(B \to \ell_1 N)/d \Omega_{{\hat p}'_N} = \Gamma(B \to \ell_1 N)/(4 \pi)$ since $B$ is a pseudoscalar, and the expression of $\Gamma(B \to \ell_1 N)$ is given in subsection II A. The nonsurvival probability $P_N$ is in the case of Eq.~(\ref{BreffnoD}) also simpler, because it (and the energy of $N$ in the lab frame, $E''_N$) depends only on the direction ${\hat p}'_N$ of $N$ in the $B$-rest frame. The expression $E''_N({\hat p}'_N)$ is given in Appendix \ref{appENpp}.

\textcolor{black}{On the other hand, in the LHCb experiment, the entire procedure described in this Section, designed for a given momentum $(p_B)_{\rm lab} \equiv p_B^{''}$ of $B$ in the laboratory frame [cf.~Eq.~(\ref{gammabetaB}) for Belle-II where $p_B=1.504$ GeV], has to be repeated for various values of momenta $p_B^{''}$. The obtained effective branching ratios then have to be averaged over these momenta $p_B^{''}$. We took into account that the lab momentum $p_B^{''}$ of the produced $B$ mesons in LHCb is distributed over a large interval, cf.~the shaded curve in Fig.~\ref{Bdistr}(a).\footnote{We thank Sheldon L. Stone (LHCb Collaboration) for providing us with the distribution, from Ref.~\cite{LHCC98004}, appearing here as Fig.~\ref{Bdistr}(a).}
\begin{figure}[htb] 
\begin{minipage}[b]{.49\linewidth}
\includegraphics[width=85mm,height=50mm]{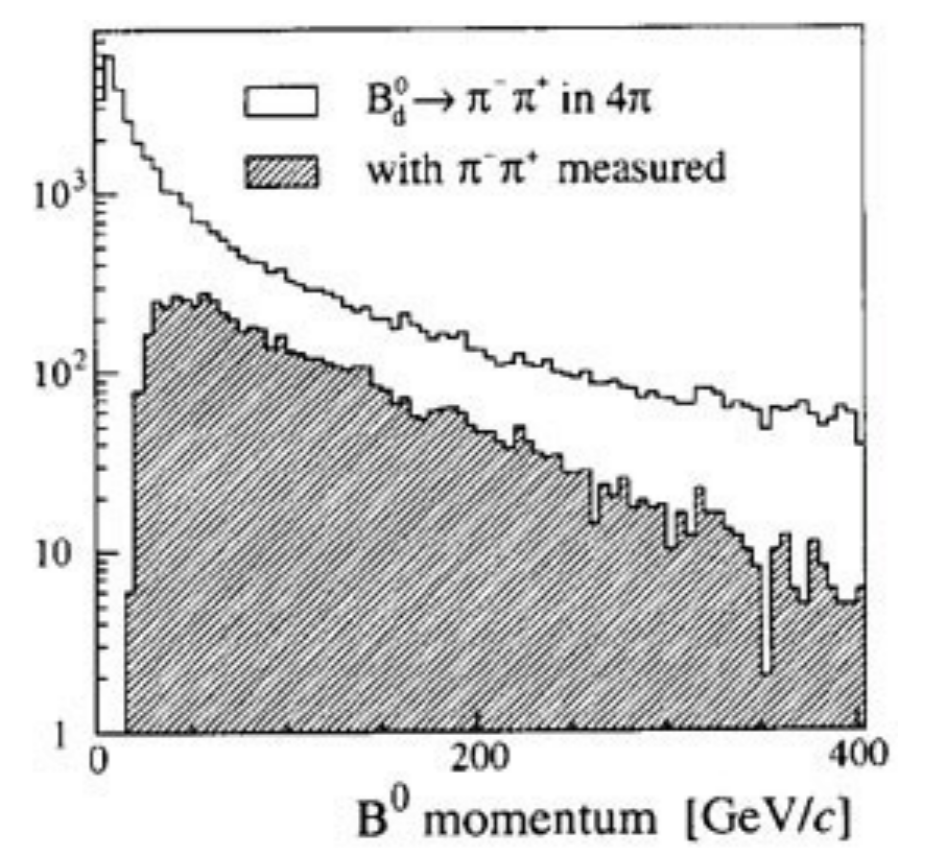}
\end{minipage}
\begin{minipage}[b]{.49\linewidth}
\includegraphics[width=75mm,height=47mm]{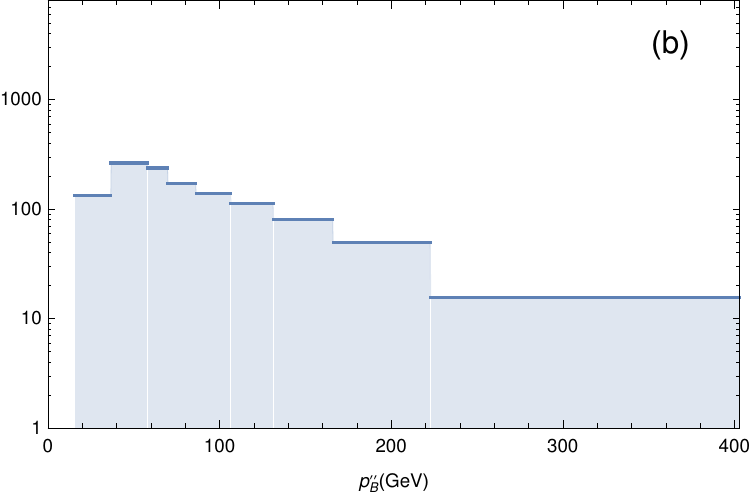}
\end{minipage} \vspace{12pt}
\caption{\textcolor{black}{\footnotesize (a) (left-hand figure) The lab momentum  ($p_B^{''}$) distribution of the produced $B^0$ mesons in LHCb \cite{LHCC98004}. We take the shaded figure as the representative case; (b) (right-hand figure) the distribution of the left-hand shaded curve in ten bins of equal weight (equal number of events).}}
\label{Bdistr}
\end{figure}
We separated this distribution in ten bins of equal weight (equal number of events), cf.~Fig.~\ref{Bdistr}(b), and calculated the results of Figs.~\ref{figUmuN2LHCb}(a)-(d) by averaging over these ten bins. For each bin, we took in our evaluations the value of the $B$ meson momentum to be such that, within the bin interval, the number of events to the left and to the right of it [according to the shaded curve of Fig.~\ref{Bdistr}(a)] are equal; e.g., in the last bin, $223 \ {\rm GeV} < p_N < 403 \ {\rm GeV}$, the average momentum value taken is $p = 273 \ {\rm GeV}$.}

\section{Numerical results for sensitivity limits on $|U_{\mu N}|^2$ 
at LHCb upgrade and Belle-II}
\label{sec:num}

We assume that in the considered decays, the produced on-shell neutrino $N$ has the available length of $L=1 \ m$ for flight within the detector,
\textcolor{black}{at Belle-II and $L=2.3 \ m$ at LHCb upgrade.}\footnote{This length $L$ is considered here to be independent of the position of the vertex where $N$ is produced and independent of the direction in which the produced $N$ travels. It can be called here the effective detector length for the neutrino $N$.
  \textcolor{black}{In the case of LHCb, the length of the Vertex Locator (VELO) is about $1 \ m$ \cite{VELO}; the effective detector length could be extended beyond that locator, to $L=2.3 \ m$ \cite{RICHdesign,Sheldon}.}} We consider that at Belle-II,
\textcolor{black}{the total number of $5 \times 10^{10}$ $B$-mesons will be produced \cite{Belle-II}, and at LHCb upgrade this number will be about $4.8 \times 10^{12}$ \cite{Sheldon}.}
We assume that there are no background events for the considered lepton number violating (LNV) decays
\textcolor{black}{
$B \to D^{(*)} \mu^{\pm} N \to D^{(*)} \mu^{\pm} \mu^{\pm} X^{\mp}$; and $B^{\pm} \to \mu^{\pm} N \to \mu^{\pm} \mu^{\pm} X^{\mp}$. Here, $X^{\pm}$ stands either for $\pi^{\pm}$ (LHCb and Belle-II), or the lepton pair $e^{\pm} \nu_e$  (Belle-II), and $B$ stands for $B^0$, ${\bar B}^0$ or $B^{\pm}$.}
In these events, we have no QED background because no $\mu^+ \mu^-$ pairs appear in the final states.

The effective branching ratios of the mentioned decay modes depend crucially on the heavy-light mixing parameter $|U_{\mu N}|^2$. The sensitivity limit on $|U_{\mu N}|^2$ at 95 \% confidence limit is obtained for $N_{\rm events}=3.09$ \cite{FC}. Therefore, the sensitivity limits on $|U_{\mu N}|^2$ are obtained by requiring
\textcolor{black}{ $\langle {\rm Br}_{\rm eff} \rangle = 3.09/(4.8 \times 10^{12})$ at LHCb upgrade, and  $\langle {\rm Br}_{\rm eff} \rangle = 3.09/(5 \times 10^{10})$ at Belle-II, where we recall that the projected total number of produced $B$ mesons at LHCb upgrade and at Belle-II is $4.8 \times 10^{12}$ and $5 \times 10^{10}$, respectively.}

\textcolor{black}{The values of $\langle {\rm Br}_{\rm eff}(B \to D^{\star} \mu \mu X) \rangle $ ($X=\pi$ or $e \nu_e$) are obtained by taking the arithmetic average of the values of ${\rm Br}_{\rm eff}$ for the four LNV decay modes: $B^- \to D^{\star 0} \mu^- \mu^- X^+$, ${\bar B}^0 \to D^{\star +} \mu^- \mu^- X^+$ and their charge conjugates. Analogously, $\langle {\rm Br}_{\rm eff}(B \to D \mu \mu X ) \rangle$ is the arithmetic average over the four analogous LNV decays as mentioned before, having now $D$ instead of $D^{\star}$. We note that the total decay widths of $B^0$ and $B^{\pm}$ differ somewhat, $\Gamma_{B^0}/\Gamma_{B^+} =1.078$ \cite{PDG2016}, and we took this into account. In our calculations we neglected, however, the small difference between the masses of $D^+$ and $D^0$ (about $5$ MeV), and between the masses of $D^{\star +}$ and $D^{\star 0}$ (about $3$ MeV); we used $m_D \approx 1.865$ GeV and $m_{D^{\star}} \approx 2.010$ GeV.}

\textcolor{black}{Further, for the LNV decays of $B$ without $D^{(*)}$ mesons, $B \to \mu \mu X$, we do not have four, but only two modes, due to the electric charge restriction: $B^{\pm} \to \mu^{\pm} \mu^{\pm} X^{\mp}$. For such decays, the average $\langle {\rm Br}_{\rm eff}(B \to \mu \mu X) \rangle$ is taken only over these two LNV modes. In these latter cases, we have to take into account that the total number of produced charged $B$ mesons is only half of the total number of produced $B$ mesons. Hence,  the sensitivity limits on $|U_{\mu N}|^2$ are obtained in these cases by requiring $\langle {\rm Br}_{\rm eff} (B \to \mu \mu X) \rangle = 3.09/(2.4 \times 10^{12})$ at LHCb upgrade, and  $\langle {\rm Br}_{\rm eff} \rangle = 3.09/(2.5 \times 10^{10})$ at Belle-II.}

\textcolor{black}{We note that the charge-conjugated versions of the decays, i.e., the decays of $B^0$ vs ${\bar B}^0$, and of $B^+$ vs $B^-$, give in general the same results. The only exception are the decays in which $D^*$ vector meson is produced. This is so because of the factor $\eta=\pm 1$ in the expression (\ref{dGdq2domdom2}), in one term there proportional to $\cos \theta_1$, which changes sign. The effect of this sign change does not entirely cancels out in the integration (\ref{Breff}) for the effective branching ratio, because the expression $E''_N(q^2;{\hat q}',{\hat p}_{1})$ in the neutrino $N$ decay probability also has dependence on $\cos \theta_1$.}

  We assume in our formulas that only the mixings $|U_{\mu N}|^2$ are nonzero; if other mixings ($|U_{e N}|^2$, $|U_{\tau N}|^2$) are nonzero, the obtained upper bounds for  $|U_{\mu N}|^2$ are in general less restrictive (higher).\footnote{
If ${\bar N}$ (and $N$) were Dirac, it would produce, e.g., a pair $\mu^+ \mu^-$ or a pair $e^+ e^-$, which have a strong QED background, and would thus not be useful. Or it could produce a pair $\mu^{\pm} e^{\mp}$; this could give important contribution, but only in the scenario where both $U_{\mu N}$ and $U_{e N}$ are nonnegligible, i.e., the scenario not considered here.}

\begin{figure}[htb] 
\begin{minipage}[b]{.49\linewidth}
  \centering\includegraphics[width=85mm]{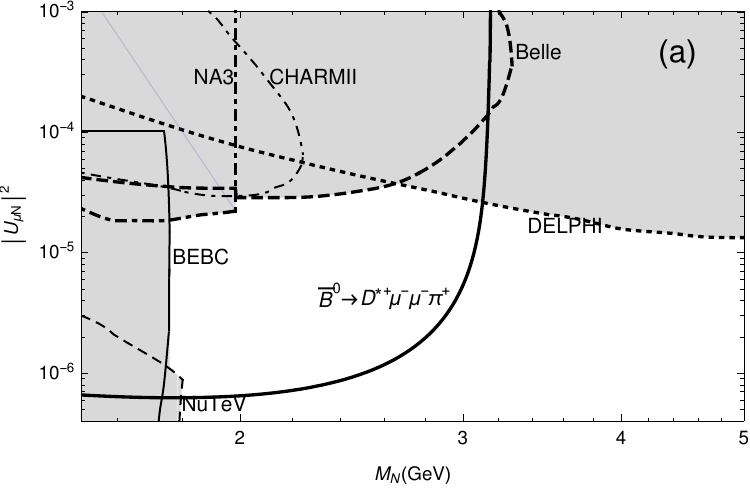}
  \end{minipage}
\begin{minipage}[b]{.49\linewidth}
  \centering\includegraphics[width=85mm]{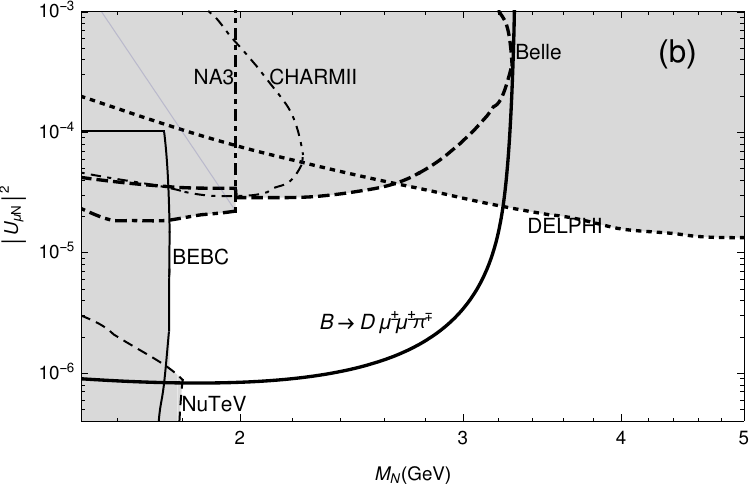}
\end{minipage}
\begin{minipage}[b]{.49\linewidth}
  \centering\includegraphics[width=85mm]{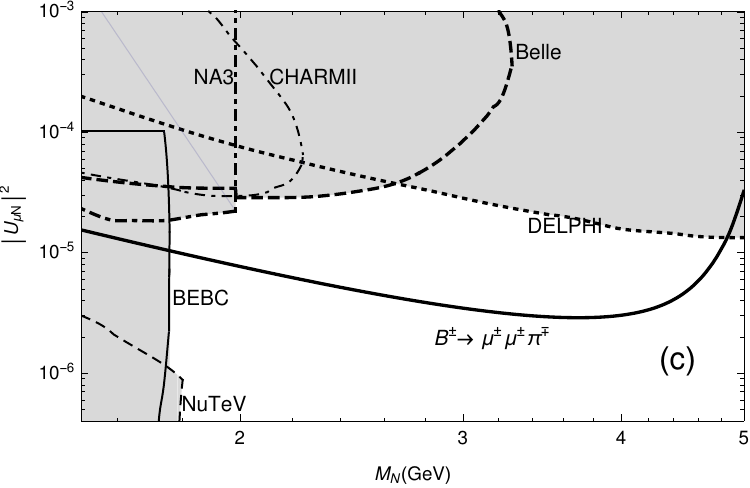}
  \end{minipage}
\begin{minipage}[b]{.49\linewidth}
  \centering\includegraphics[width=85mm]{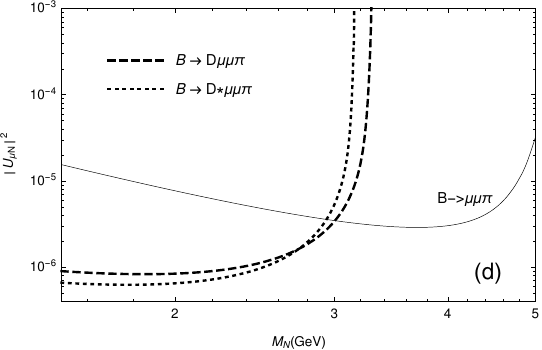}
\end{minipage}
\caption{\textcolor{black}{\footnotesize (a) The sensitivity limits on $|U_{\mu N}|^2$ at LHCb upgrade, as solid lines, from LNV decays $B \to D^{*} \mu^{\pm} N \to D^{*} \mu^{\pm} \mu^{\pm} \pi^{\mp}$; for comparison, the present bounds from various experiments are included, giving the grey region of exclusion. (b) As (a), but for the decays $B \to D \mu^{\pm} N \to D \mu^{\pm} \mu^{\pm} \pi^{\mp}$. (c) As (a), but for the decays $B^{\pm} \to \mu^{\pm} N \to \mu^{\pm} \mu^{\pm} \pi^{\mp}$. (d) Comparison of the prospective LHCb sensitivity limits for the three decays. The effective detector length is taken $L=2.3 \ m$, and the expected total number of produced $B$ meson pairs $N=4.8 \times 10^{12}$.}}
\label{figUmuN2LHCb}
\end{figure}
\textcolor{black}{The results for the decays with $\pi^{\pm}$ in the final state, for LHCb upgrade, are given in Figs.~\ref{figUmuN2LHCb}(a)-(d).
In Figs.~\ref{figUmuN2LHCb}(a)-(c), the present direct experimental bounds are included for comparison, along with our results - the obtained prospective sensitivity limits for LHCb upgrade. Fig.~\ref{figUmuN2LHCb}(d) shows the LHCb sensitivity limits for the three considered decays, for mutual comparisons. Further, we note that the decay modes $B \to (D^{(*)}) \mu^{\pm} \mu^{\pm} e^{\mp} \nu_e$ cannot be detected at LHCb.}

\begin{figure}[htb] 
\begin{minipage}[b]{.49\linewidth}
  \centering\includegraphics[width=85mm]{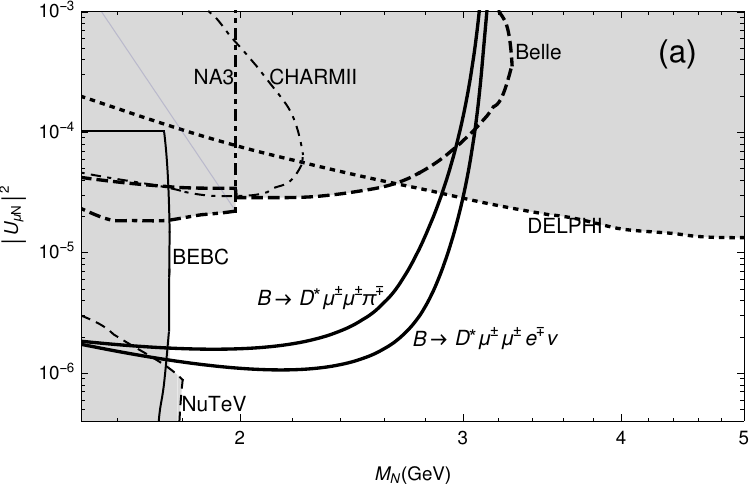}
  \end{minipage}
\begin{minipage}[b]{.49\linewidth}
  \centering\includegraphics[width=85mm]{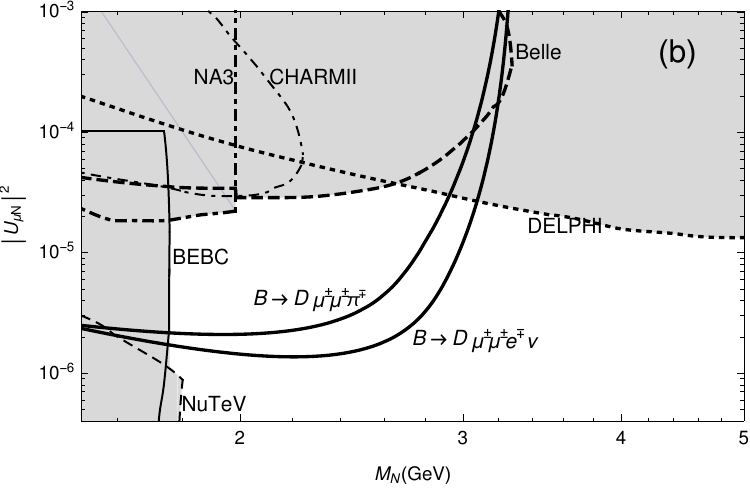}
\end{minipage}
\begin{minipage}[b]{.49\linewidth}
  \centering\includegraphics[width=85mm]{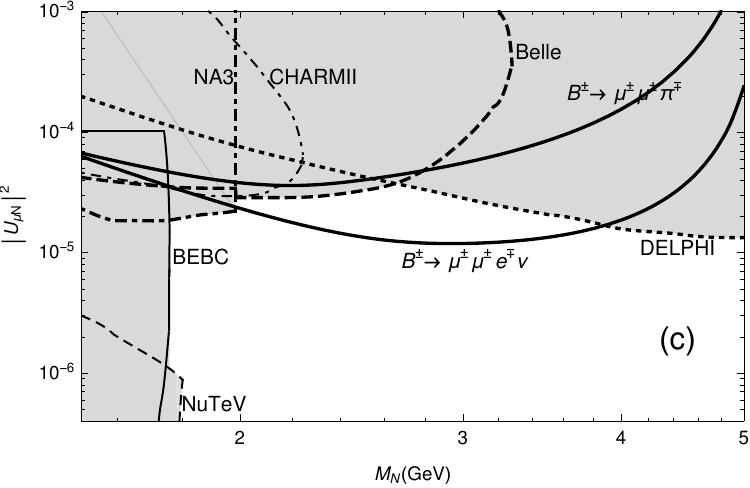}
  \end{minipage}
\begin{minipage}[b]{.49\linewidth}
  \centering\includegraphics[width=85mm]{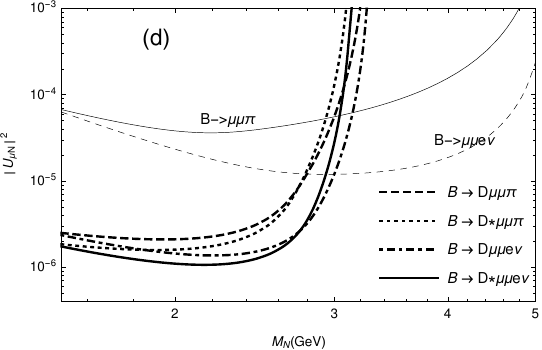}
\end{minipage}
\caption{\footnotesize (a) The future Belle-II sensitivity limits on $|U_{\mu N}|^2$, as solid lines, from LNV decays $B \to D^{*} \mu^{\pm} N \to D^{*} \mu^{\pm} \mu^{\pm} X^{\mp}$ at Belle-II, where  $X^{\mp}=\pi^{\mp}$ or $X^{\mp}=e^{\mp} \nu_e$; included are also the present bounds from various experiments, resulting in the grey region of exclusion. (b) The same, but for the decays $B^ \to D \mu^{\pm} N \to D \mu^{\pm} \mu^{\pm} X^{\mp}$. (c) The same, but for the decays $B^{\pm} \to \mu^{\pm} N \to \mu^{\pm} \mu^{\pm} X^{\mp}$. (d) Comparison of the prospective Belle-II sensitivity limits for the three mentioned pairs of decays.  The effective detector length is taken $L=1 \ m$, and the expected total number of produced $B$ meson pairs $N=5 \times 10^{10}$.}
\label{figUmuN2}
\end{figure}
The results for the considered decays
\textcolor{black}{at Belle-II,  either with $\pi^{\pm}$ or with $e^{\pm} \nu_e$ in the final state},
are given in Figs.~\ref{figUmuN2}(a)-(d). In Figs.~\ref{figUmuN2}(a)-(c), the present experimental bounds are included for comparison. In Fig.~\ref{figUmuN2}(d), the prospective Belle-II sensitivity limits for all the six considered decays are presented, for mutual comparisons.

\section{Discussions and conclusions}
\label{sec:concl}

From Figures \ref{figUmuN2LHCb} and \ref{figUmuN2}, we can see that the decays where $D^{*}$ and $D$ are produced give quite strong new sensitivity limits on $|U_{\mu N}|^2$ in the mass interval $1.75 \ {\rm GeV} < M_N < 3 \ {\rm GeV}$. This is a reflection of the fact that the presence of $D^{(*)}$ mesons leads to a significantly weaker CKM suppression in the decay rates, because $|V_{cb}|^2 \approx 10^2 |V_{ub}|^2$. However, when $M_N > 3 \ {\rm GeV}$, such decays are kinematically suppressed, and then only the (CKM-suppressed) decays $B \to \mu \mu X$ give useful sensitivity limits, as seen in Figs.~\ref{figUmuN2LHCb}(c), (d) and Figs.~\ref{figUmuN2}(c), (d). Further, we see in Figs.~\ref{figUmuN2LHCb} that in general the sensitivity limits are more restrictive (lower) when $X=e \nu$ than when $X=\pi$.

\textcolor{black}{Comparing Figs.~\ref{figUmuN2LHCb} with Figs.~\ref{figUmuN2}, we can see that the decays $B \to (D^{(*)}) \mu^{\pm} \mu^{\pm} \pi^{\mp}$, which can be measured at both LHCb and Belle-II experiments, give more stringent (lower) sensitivity limits on $|U_{\mu N}|^2$ at LHCb upgrade experiment. This is so primarily because the expected number of produced $B$ mesons at LHCb upgrade ($4.8 \times 10^{12}$) is by two orders of magnitude larger than the number at Belle-II ($5 \times 10^{10}$). Yet another factor contributing to the more stringent bounds is the effective detector length, which is assumed to be larger at LHCb upgrade ($L=2.3 \ m$  vs $L=1 \ m$ at Belle-II). The difference between the two sets of the sensitivity limits is somewhat reduced by the fact that the lab energy of the produced $B$ mesons in LHCb is significantly higher than in Belle-II; as a consequence, the produced on-shell $N$ neutrinos move in the LHCb case faster and are thus less likely to decay within the detector. If, on the other hand, the acceptance factors decrease the effective number $N$ of produced $B$ mesons, or if the effective detector length $L$ turns out to be smaller, the sensitivity limits for $|U_{\mu N}|^2$ go up, in general as approximately proportional to $1/\sqrt{N L}$ for not very heavy neutrinos ($M_N < 3$ GeV).}

\textcolor{black}{This approximate proportionality comes from the following behavior. For the values of $|U_{\mu N}|^2$ which are of the order of magnitude of the presented upper bounds, we have at $M_N \lesssim 2.5$ GeV small $N$-decay probabilities, $P_N \ll 1$, and therefore our expressions imply in such a case the approximate proportionality ${\rm Br}_{\rm eff} \propto |U_{\mu N}|^4 L$. However, for $M_N \gtrsim 4.5$ GeV we have $P_N \approx 1$ and thus the approximate proportionality ${\rm Br}_{\rm eff} \propto  |U_{\mu N}|^2$ (and $L$-independent). We verified these approximate proportionalities also numerically with our expressions. Approximate $L$-independence of ${\rm Br}_{\rm eff}$ occurs already at $M_N \gtrsim 3$ GeV.}

In Ref.~\cite{AsIsh}, a similar analysis was made for the decay $B^+ \to \mu^+ N \to \mu^+ \mu^- \pi^-$ at Belle-II, where the same total number of $B$ meson pairs was assumed as here, $5 \times 10^{10}$. They obtained lower, i.e., more restrictive sensitivity limits on $|U_{\mu N}|^2$ than we do for this decay for Belle-II. The reason for the difference cannot be the fact that they did not take into account the movement of $B$-mesons in the lab frame (this effect changes the sensitivity limits only weakly). The reason for the difference lies possibly in the evaluated values of the total decay width $\Gamma_N$ as a function of $M_N$. We evaluated this decay width according to the formulas and Figures in Appendix \ref{appNall}, based on Refs.~\cite{Atre,HKS}, and we applied those evaluations in Refs.~\cite{CKZ2,symm}.

The experimental bounds on $|U_{\mu N}|^2$ presented in Figs.~\ref{figUmuN2LHCb}(a)-(c)  and Figs.~\ref{figUmuN2}(a)-(c) are from various experiments: DELPHI \cite{DELPHI},  BEBC \cite{BEBC}, NuTeV \cite{NuTeV}, NA3 \cite{NA3}, CHARM II \cite{CHARMII}, and Belle \cite{BelleUB}.

On the basis of the obtained results, Figs.~\ref{figUmuN2LHCb} and \ref{figUmuN2}, we conclude that the LHCb upgrade and Belle-II experiments have the potential to either find a new heavy Majorana neutrino $N$, or to improve significantly the sensitivity limits (upper bounds) on the heavy-light mixing parameter $|U_{\mu N}|^2$, particularly in the mass range $1.75 \ {\rm GeV} < M_N < 3 \ {\rm GeV}$ where the LNV decays of $B$ mesons involving $D$ or $D^{*}$ mesons and an on-shell neutrino $N$ are possible.

If $N$ is not Majorana but Dirac particle, then clear sensitivity limits cannot be obtained for $|U_{\mu N}|^2$, but rather for the product $|U_{e N} U_{\mu N}|$; this is a less attractive possibility, principally because the present upper bounds for $|U_{e N}|^2$ in the mentioned mass range, coming from the neutrinoless double beta decay experimental data \cite{Kova}, are more restrictive (lower) than those for $|U_{\mu N}|^2$.

\section*{Acknowledgments}
\noindent
The work of C.S.K. was supported by the NRF grant funded by the Korean government of the MEST (No. 2016R1D1A1A02936965).
\textcolor{black}{We thank Y.J.~Kwon and Sheldon L.~Stone for providing us with valuable information on Belle-II and LHCb upgrade experiments, respectively.}

\appendix

\section{Total decay width of neutrino $N$}
\label{appNall}

We summarize here the formulas needed for evaluation of the total decay width of a massive sterile neutrino $N$, $\Gamma(N \to {\rm all})$.

The formulas for the widths for leptonic decays and semileptonic decay are given in Ref.~\cite{Atre} (Appendix C there), for $M_N \lesssim 1$ GeV. For higher masses $M_N$, the calculation of the semileptonic decay widths cannot be performed in this way because not all the resonances are known. For such higher masses, the decay widths for semileptonic decays were calculated in Refs.~\cite{HKS,GKS} by the inclusive approach based on duality. In this approach, the various (pseudoscalar and vector) meson channels were calculated by quark-antiquark channels. This was applied for $M_N \geq M_{\eta^{'}} \approx 0.958$ GeV. Below we write the expressions given in Ref.~\cite{HKS} for the decay width channels. In some of these formulas, twice the decay width appears [$2 \Gamma(N \to \ldots)$], where the factor two is applied if $N$ is a Majorana neutrino, and factor one if it is Dirac neutrino. This is so because when Majorana neutrino decays to charged particles, the decay in charge conjugate channel is equally possible; this is not possible if $N$ is Dirac~particle.

The leptonic decays are
\bes
\label{GNlept}
\bea
2 \Gamma(N \to \ell^- \ell^{'+} \nu_{\ell^{'}}) & = &
|U_{\ell N}|^2 \frac{G_F^2}{96 \pi^3} M_N^5 I_1(y_{\ell},0, y_{\ell^{'}})
(1 - \delta_{\ell \ell^{'}} ) \ ,
\label{GNlepta}
\\
\Gamma(N \to \nu_{\ell} \ell^{'-} \ell^{'+}) & = &
|U_{\ell N}|^2 \frac{G_F^2}{96 \pi^3} M_N^5
{\big [}
(g_L^{(\rm lept)} g_R^{(\rm lept)}
+ \delta_{\ell \ell^{'}} g_R^{(\rm lept)}) I_2(0,y_{\ell^{'}},y_{\ell^{'}})
\nonumber\\
&&
+ \left( (g_L^{(\rm lept)})^2 + (g_R^{(\rm lept)})^2
+  \delta_{\ell \ell^{'}}
(1 + 2 g_L^{(\rm lept)}) \right) I_1(0,y_{\ell^{'}},y_{\ell^{'}})
{\big ]}
\label{GNleptb}
\\
\sum_{\nu_{\ell}} \sum_{\nu^{'}} \Gamma(N \to \nu_{\ell} \nu^{'} {\bar \nu}^{'})
& = &
\sum_{{\ell}} |U_{\ell N}|^2 \frac{G_F^2}{96 \pi^3} M_N^5 \
\label{GNleptc}
\eea
\ees
The factor $2$  is included in Eq.~(\ref{GNlepta}) when $N$ is Majorana, because in such a case both decays, $N \to \ell^- \ell^{'+} \nu_{\ell^{'}}$ and $N \to \ell^+ \ell^{'-} \nu_{\ell^{'}}$ are contributing ($\ell \not= \ell^{'}$).

Further, the following semileptonic decays contribute when $M_N < M_{\eta^{'}} \approx 0.968$ GeV, involving pseudoscalar ($P$) and vector ($V$) mesons:
\bes
\label{GNmes}
\bea
2 \Gamma(N \to  \ell^- P^+) & = &
|U_{\ell N}|^2 \frac{G_F^2}{8 \pi} M_N^3 f_P^2 |V_P|^2 F_P(y_{\ell}, y_P) \
\label{GNMesPch}
\\
\Gamma(N \to  \nu_{\ell} P^0) & = &
|U_{\ell N}|^2 \frac{G_F^2}{64 \pi} M_N^3 f_P^2 (1 - y_P^2)^2 \
\label{GNMesP0}
\\
2 \Gamma(N \to  \ell^- V^+) & = &
|U_{\ell N}|^2 \frac{G_F^2}{8 \pi} M_N^3 f_V^2 |V_V|^2 F_V(y_{\ell}, y_V) \
\label{GNMesVch}
\\
\Gamma(N \to  \nu_{\ell} V^0) & = &
|U_{\ell N}|^2 \frac{G_F^2}{2 \pi} M_N^3 f_V^2 \kappa_V^2 (1 - y_V^2)^2
(1 + 2 y_V^2) .
\label{GNMesV0}
\eea
\ees
Again, the factor $2$ appears in the charged meson channels if $N$ is Majorana.
The factors $V_P$ and $V_V$ appearing in the above expressions stand for the CKM
matrix elements of the valence quarks of the mesons. Ths constants $f_P$ and $f_V$ are the corresponding decay constants of these mesons. Their values are given in Table 1 of Ref.~\cite{HKS}.

The contributing pseudoscalar mesons here are:
$P^{\pm} = \pi^{\pm}, K^{\pm}$; $P^0 = \pi^0, K^0, {\bar K}^0, \eta$.
The contributing vector mesons here are:
$V^{\pm} = \rho^{\pm}, K^{* \pm}$; $V^0= \rho^0, \omega, K^{*0}, {\bar K}^{*0}$.

On the other hand, for higher mass $M_N \geq M_{\eta^{'}}$ ($=$0.9578 GeV), the quark-hadron duality is used and the sum of the widths of the semileptonic decay modes are represented by the following widths into quark-antiquark decay modes \cite{HKS}:
{\small
\bes
\label{GNquark}
\bea
2 \Gamma(N \to  \ell^- U {\bar D}) & = &
|U_{\ell N}|^2 \frac{G_F^2}{32 \pi^3} M_N^5 |V_{UD}|^2 I_1(y_{\ell},y_U,y_D) \
\label{GNquarka}
\\
\Gamma(N \to  \nu_{\ell} q {\bar q}) & = &
|U_{\ell N}|^2 \frac{G_F^2}{32 \pi^3} M_N^5
\left[
g_L^{(q)} g_R^{(q)} I_2(0,y_q,y_q)\!+\!
\left( (g_L^{(q)})^2\!+\!(g_R^{(q)})^2 \right) I_1(0,y_q,y_q)
\right] \
\label{GNquarkb}
\eea
\ees}
\vspace{-6pt}

In all the formulas (\ref{GNlept})--(\ref{GNquark}), the notations
\be
y_{Y} \equiv M_Y/M_N \qquad (Y = \ell, \nu_{\ell},  P, V, q)
\label{GNnot}
\ee
are used. We denoted in Eq.~(\ref{GNquark}): $U=u,c$; $D=d,s,b$; $q=u,d,c,s,b$. The used values of the quark masses in our evaluations are:
$M_u=M_d = 3.5$ MeV; $M_s=105$ MeV; $M_c=1.27$ GeV; $M_b=4.2$ GeV.

As mentioned earlier, in the evaluation of the total decay width $\Gamma_N$,
if $N$ is Majorana we add the expressions (\ref{GNquarka}) and (\ref{GNquarkb}); if $N$ is Dirac, the expressions should be added, but with the expressions (\ref{GNquarka}) multiplied by $1/2$. The same is valid in the case when we sum the expressions (\ref{GNlept}) and (\ref{GNmes}).

In Eqs.~(\ref{GNleptb}) and (\ref{GNquarkb}), the following SM neutral current couplings appear:
\bes
\label{NCc}
\bea
g_L^{(\rm lept)} &=& - \frac{1}{2} + \sin^2 \theta_W \ ,
\quad
g_R^{(\rm lept)} =\sin^2 \theta_W \
\label{NCcl}
\\
g_L^{(U)} & = & \frac{1}{2} - \frac{2}{3} \sin^2 \theta_W \ ,
\quad
g_R^{(U)} = - \frac{2}{3} \sin^2 \theta_W \
\label{NCcU}
\\
g_L^{(D)} & = & - \frac{1}{2} + \frac{1}{3} \sin^2 \theta_W \ ,
\quad
g_R^{(U)} =  \frac{1}{3} \sin^2 \theta_W \
\label{NCcD}
\eea
\ees

In Eq.~(\ref{GNMesV0}), the neutral current couplings $\kappa_V$ (for the neutral vector mesons) are
\bes
\label{kV}
\bea
\kappa_V & = & \frac{1}{3} \sin^2 \theta_W  \quad (V=\rho^0, \omega) \
\label{kVa}
\\
\kappa_V & = & - \frac{1}{4} +\frac{1}{3} \sin^2 \theta_W
\quad (V=K^{*0}, {\bar K}^{*0}) \
\label{kVb}
\eea
\ees

Further, in the above expressions,  the following expressions $I_1$, $I_2$, $F_P$ and $F_V$ were used:
\bes
\label{kinex}
\bea
I_1(x,y,z) & = & 12 \int_{(x+y)^2}^{(1-z)^2} \; \frac{ds}{s}
(s - x^2 - y^2) (1 + z^2 -s)
\lambda^{1/2}(s,x^2,y^2) \lambda^{1/2}(1,s,z^2) \
\label{I1}
\\
I_2(x,y,z) & = & 24 y z \int_{(y+z)^2}^{(1-x)^2} \; \frac{ds}{s}
(1 + x^2 - s)
\lambda^{1/2}(s,y^2,z^2) \lambda^{1/2}(1,s,x^2) \
\label{I2}
\\
F_P(x,y) & = & \lambda^{1/2}(1,x^2,y^2) \left[(1 + x^2)(1 + x^2-y^2) - 4 x^2
\right] \
\label{FP}
\\
F_V(x,y) & = & \lambda^{1/2}(1,x^2,y^2) \left[(1 - x^2)^2 + (1 + x^2) y^2 - 2 y^4 \right] .
\label{FV}
\eea
\ees
Here, the $\lambda^{1/2}$ function is given in Eq.~(\ref{lam}).

\begin{figure}[b] 
\begin{minipage}[b]{.49\linewidth}
\includegraphics[width=85mm]{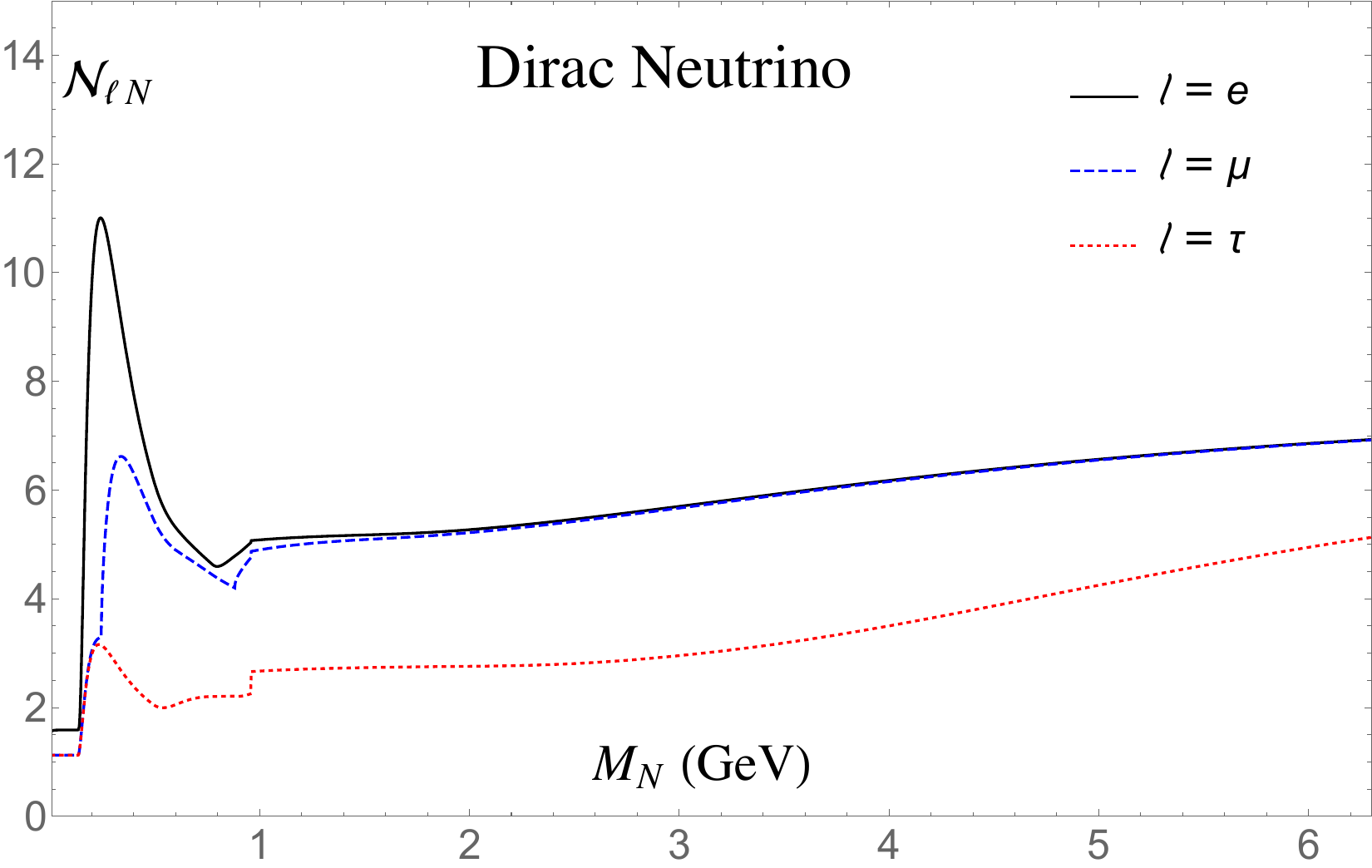}
\end{minipage}
\begin{minipage}[b]{.49\linewidth}
\includegraphics[width=85mm]{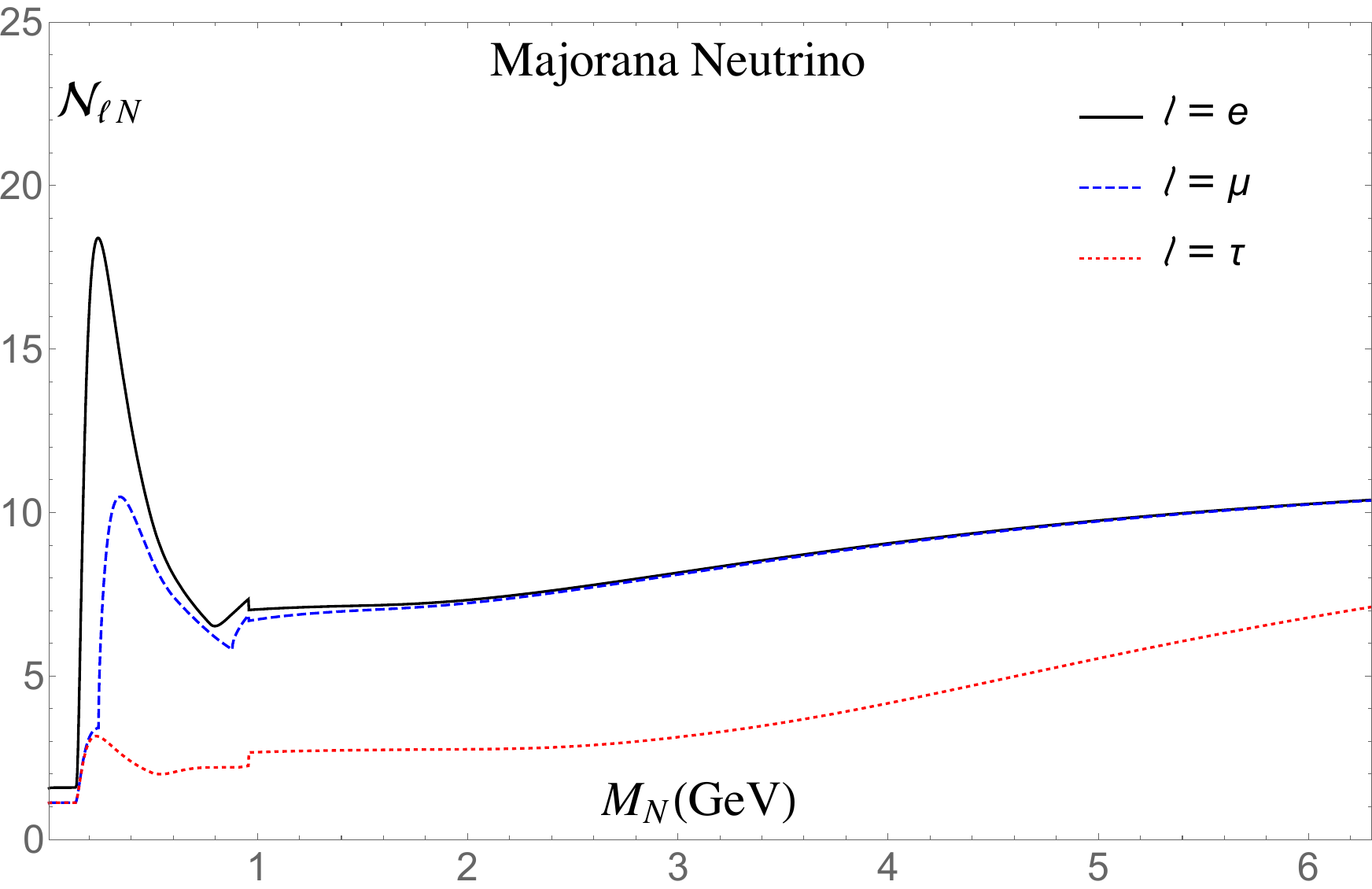}
\end{minipage} \vspace{12pt}
\caption{The coefficients ${\cal N}_{\ell N}(M_N)$ ($\ell = e, \mu, \tau$) appearing in Eqs.~(\ref{GNwidth})-(\ref{calK}), as a function of the mass of the heavy sterile neutrino $N$. When $N$ is Dirac, the left-hand figure applies; when it is Majorana, the right-hand figure applies.}
\label{FigcNellN}
\end{figure}

All these formulas then give the total decay width $\Gamma(N \to {\rm all})$ as a function of $M_N$. This total decay width can be written in the following form:
\begin{equation}
\Gamma_{N} = \K \; {\overline \Gamma}_N(M_{N}) \ .
\label{GNwidth}
\end{equation}
The corresponding canonical (i.e., without the heavy-light mixing factors) decay width expression is
\begin{equation}
 {\overline \Gamma}_N(M_{N}) \equiv \frac{G_F^2 M_{N}^5}{96 \pi^3} \ .
\label{barGN}
\end{equation}
The factor $\K$ in Eq.~({GNwidth}) contains the dependence on the heavy-light mixing factors, and it has the form
\begin{equation}
\K(M_{N}) \equiv \K = {\cal N}_{e N} \; |U_{e N}|^2 + {\cal N}_{\mu N} \; |U_{\mu N}|^2 + {\cal N}_{\tau N} \; |U_{\tau N}|^2  \
\label{calK}
\end{equation}
The dimensionless coefficients ${\cal N}_{\ell N}(M_N)$ here are numbers $\sim 1$-$10$ which are functions of the mass $M_N$, and they are determined by the above formulas given in this Appendix. We present in Figs.~\ref{FigcNellN} the resulting coefficients ${\cal N}_{\ell N}(M_N)$ as a function of neutrino mass $M_N$, for the case of Dirac and Majorana neutrino $N$. The figures are from Ref.~\cite{CKZ2} for Majorana $N$, and \cite{symm} for Dirac $N$.

It is interesting to notice a small kink in the curves of Figs.~\ref{FigcNellN} at $M_N=M_{\eta^{'}}$ ($=$0.9578 GeV). The kink is there because at  $M_N \geq M_{\eta^{'}}$ the use of quark-hadron duality is made, i.e., we replace the semileptonic decay channel contributions by those of the quark-antiquark channel. As a consequence, we can conclude that the quark-hadron duality works well at $M_N \geq M_{\eta^{'}}$. A partial exception is  the case $\ell = \tau$ because $\tau$ lepton has a large mass.

\section{Lorentz factors of on-shell $N$ in laboratory frame}
\label{appENpp}

In this Appendix we calculate the energy $E''_N$ of the produced heavy neutrino $N$ in the laboratory frame $\Sigma''$ [the rest frame of $\Upsilon(4S)$] in the reaction $B \to D^{(*)} \ell_1 N$, cf.~Sec.~\ref{sec:PN}. We recall that our notations are: $\Sigma$ is the rest frame of the virtual $W^{*}$ (i.e., of the $\ell_1$-$N$ pair); $\Sigma'$ is the rest frame of the $B$ meson; and $\Sigma''$ is the laboratory frame.

  As explained in Sec.~\ref{sec:PN}, the velocity of the produced mesons $B$ in the laboratory ($\Sigma''$) frame, ${\vec \beta}_B$, is (practically) the same as the velocity of $\Upsilon(4S)$ there, Eqs.~(\ref{gammabetaU})-(\ref{gammabetaB}). The momentum $p_N$ transforms between the $\Sigma''$ (lab) frame and the $\Sigma'$ (B-rest) frame in the following way:
\bes
\label{SppSp}
\bea
E''_N & = & \gamma_B \left( E'_N + \beta_B ({\vec {p'_N}} \cdot {\hat \beta}_B) \right),
\label{SppSpa}
\\
({\vec {p''_N}} \cdot {\hat \beta}_B) & = & \gamma_B \left(   ({\vec {p'_N}} \cdot {\hat \beta}_B) + \beta_B E'_N \right),
\label{SppSpb}
\\
( {\vec {p''_N}} )_{\perp} &=&  ( {\vec {p'_N}} )_{\perp},
\label{SppSpc}
\eea
\ees
where in the last line $(\ldots)_{\perp}$ denotes the component of the vector perpendicular to ${\hat \beta_B} \equiv {\hat z}^{''}$, i.e., perpendicular to the direction of movement of $B$ [$\leftrightarrow$ of $\Upsilon(4S)$] in the lab frame $\Sigma''$.\footnote{Strictly speaking, we should use the notation ${\vec {\beta''_B}}$ for the velocity of $B$ meson in the lab, but we prefer the simplified notation  ${\vec \beta}_B$ for this vector.}

\begin{figure}[t]
\centering\includegraphics[width=60mm]{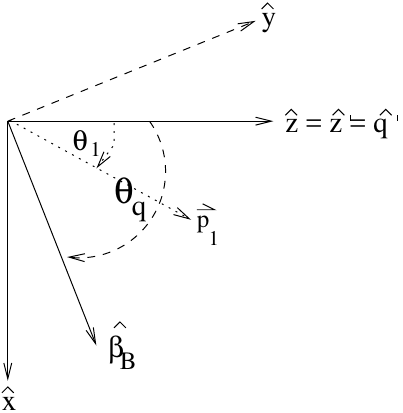}
\caption{The 3-dimensional coordinate system in the $\Sigma$ frame (the rest frame of $W^*$=$\ell_1$-$N$). The spherical coordinates of vector ${\hat \beta}_B$ in this system are: ${\hat \beta}_B=(\sin \theta_q,0, \cos \theta_q)$; and of ${\vec p}_1$ are: ${\vec p}_1 = |{\vec p}_1| (\sin \theta_1 \cos \phi_1, \sin \theta_1 \sin \phi_1 , \cos \theta_1)$; ${\hat q}'$ is the direction of the 3-momentum of $W^*$ in the frame $\Sigma'$ ($B$-rest frame); ${\hat \beta}_B$ is the direction of the 3-momentum of $B$ meson in the frame $\Sigma''$ (lab frame).}
\label{Figthqth1}
\end{figure}

The momentum $p_N$ transforms between the $B$ rest frame $\Sigma'$ and the $W^{*}$ rest frame $\Sigma$ in the following way:
\bes
\label{SpS}
\bea
E'_N & = & \gamma_W(q^2) \left( E_N(q^2) - \beta_W(q^2) |{\vec p}_N(q^2)| \cos \theta_1 \right),
\label{SpSa}
\\
({\vec {p'_N}} \cdot {\hat q}') & = & \gamma_W(q^2) \left(   -  |{\vec p}_N(q^2)| \cos \theta_1 + \beta_W(q^2) E_N(q^2) \right).
\label{SpSb}
\eea
\ees
Here, $\theta_1$ is the angle bewteen ${\hat q}' \equiv {\hat z}$ and ${\vec p}_1$ of $\ell_1$ in the $\Sigma$ frame of $\ell_1$-$N$. The corresponding quantities in the $\Sigma$ frame, as a function of the squared invariant mass of $W^{*}$, $Q^2$, are
\bes
\label{ENpNq}
\bea
E_N & = & \frac{1}{2 \sqrt{q^2}} (q^2 + M_N^2 - M_1^2),
\label{EN}
\\
|{\vec p}_N| = |{\vec p}_1| &=& \frac{1}{2} \sqrt{q^2} \lambda^{1/2} \left( 1, \frac{M_1^2}{q^2}, \frac{M_N^2}{q^2} \right),
\label{vecpNp1}
\eea
\ees
the Lorentz factors for the transition between $\Sigma'$ and $\Sigma$ are
\be
\gamma_W(q^2) = \left( 1 + \frac{|{\vec {q'}}|^2}{q^2} \right)^{1/2}, \qquad
\beta_W(q^2) = \left( \frac{q^2}{|{\vec {q'}}|^2} + 1 \right)^{-1/2},
\label{gWbW}
\ee
where the magnitude $|{\vec {q'}}|$ of the $3$-momentum of $W^{*}$ in $\Sigma'$ ($B$-rest frame) is
\be
|{\vec {q'}}| = \frac{1}{2} M_B \lambda^{1/2} \left( 1, \frac{ M_{\Dst}^2}{M_B^2}, \frac{q^2}{M_B^2} \right).
\label{vecq}
\ee
In order to combine all these relations Eqs.~(\ref{SppSp})-(\ref{vecq}) to obtain $E''_N$ as a function of $q^2$, ${\hat q}'$ and ${\hat p}_1$, we must express the $B$-meson velocity direction ${\hat \beta}_B$ in a 3-dimensional coordinate system in $\Sigma$. We introduce such a system in the following way: ${\hat z}$ is defined as ${\hat z}={\hat {z'}} = {\hat q}'$, i.e., the direction of $W^*$ in the $B$-rest frame ($\Sigma'$). Then the vectors ${\hat q}'$ and ${\hat \beta}_B$ define a plane, the angle between ${\hat q}'$($={\hat z}$) and ${\hat \beta}_B$ is $\theta_q$  ($0 \leq \theta_q \leq \pi$), and the axis ${\hat x}$ in this plane is such that $({\hat \beta}_B)_x = \sin \theta_q$ ($>0$). We recall that ${\hat {\beta}_B}$ is the direction vector of $B$ in $\Sigma''$ (lab) frame. The axis ${\hat y}$ is then obtained in the usual way, ${\hat y} = {\hat z} \times {\hat x}$, cf.~Fig.~\ref{Figthqth1}.
As a result, we have
\bes
\label{thqth1}
\bea
{\hat \beta}_B & = & \sin \theta_q {\hat x} + \cos \theta_q {\hat q}'
\label{betBthq}
\\
\Rightarrow \;
({\vec {p'_N}} \cdot {\hat \beta}_B) & = & ( {\vec {p'_N}} \cdot {\hat q}' ) \cos \theta_q + ({\vec {p'_N}} \cdot {\hat x}) \sin \theta_q .
\label{ppNbetB}
\eea
\ees
We can now take into account that ${\vec {p'_N}} \cdot {\hat x} = {\vec {p_N}} \cdot {\hat x}$, because these are components perpendicular to the boost direction ${\hat q}'={\hat z}$ between $\Sigma'$ and $\Sigma$. Since in $\Sigma$ we have ${\vec p}_1 = - {\vec p_N}$, we thus have
\be
{\vec {p'_N}} \cdot {\hat x} = {\vec {p_N}} \cdot {\hat x} = - {\vec p_1} \cdot {\hat x} = - |{\vec p_1}| \sin \theta_1 \cos \phi_1 = - |{\vec p_N}| \sin \theta_1 \cos \phi_1,
\label{ppNx}
\ee
where $\theta_1$ and $\phi_1$ are the spherical coordinates of ${\vec p}_1$ in $\Sigma$ ($0 \leq \theta_1 \leq \pi$; $0 \leq \phi_1 < 2 \pi$), cf.~Fig.~\ref{Figthqth1}. Substitution of Eq.~(\ref{ppNx}) into Eq.~(\ref{ppNbetB}), and taking into account the relation (\ref{SpSb}) then gives
\be
({\vec {p'_N}} \cdot {\hat \beta}_B) = \left[ \gamma_W(q^2) \left( - |{\vec p}_N(q^2)| \cos \theta_1 + \beta_W(q^2) E_N(q^2) \right) \cos \theta_q - |{\vec p}_N(q^2)| \sin \theta_1 \cos \phi_1 \sin \theta_q \right].
\label{ppNbetBfin}
\ee
Using this expression, and the expression for $E'_N$ of Eq.~(\ref{SpSa}), in the Lorentz transformation (\ref{SppSpa}), we finally obtain the energy $E''_N$ of the $N$ neutrino in the lab frame in terms of $q^2$, ${\hat q}'$ (i.e., $\theta_q$) and ${\hat p}_1$ (i.e., $\theta_1$ and $\phi_1$)
\bea
E''_N(q^2; \theta_q; \theta_1, \phi_1) &=& \gamma_B {\Big \{}
\gamma_W(q^2) \left( E_N(q^2) - \beta_W(q^2) |{\vec p}_N(q^2)| \cos \theta_1 \right)
\nonumber\\
&&
+ \beta_B \left[ \gamma_W(q^2) \left( - |{\vec p}_N(q^2)| \cos \theta_1 + \beta_W(q^2) E_N(q^2) \right) \cos \theta_q - |{\vec p}_N(q^2)|\sin \theta_1 \cos \phi_1 \sin \theta_q \right] {\Big \}}.
\label{ENppfin}
\eea
Here, the expressions $\gamma_W(q^2)$ and $\beta_W(q^2)$ are given in Eq.~(\ref{gWbW}), and the expressions for $E_N(q^2)$, $|{\vec p}_N(q^2)|$  and $|{\vec {q'}}|$ are given in Eqs.~(\ref{ENpNq}) and (\ref{vecq}).

For the decay $B \to D \ell N$ the same expressions apply, with the only difference that instead of $M_{D^*}$ we have $M_D$.

However, when the decay is without $D^{(*)}$, namely $B \to \ell_1 N$, the expression for $E''_N$ gets simplified significantly, and has only dependence on the direction ${\hat {p'}}_N$ of the $N$ neutrino in the B-rest frame ($\Sigma'$)
\be
\label{EppnoD}
E''_N  =  \gamma_B ( E'_N + \cos \theta_N \beta_B |{\vec {p'_N}}| ),
\ee
where $\theta_N$ is the angle between ${\hat \beta}_B$ and ${\vec {p'_N}}$ in the $B$-rest frame ($\Sigma'$), and we have
\bes
\label{Np}
\bea
E'_N &=& \frac{M_B^2 + M_N^2 - M_1^2}{2 M_B},
\label{ENp}
\\
|{\vec {p'_N}}| & = & \frac{1}{2} M_B \lambda^{1/2} \left( 1, \frac{M_1^2}{M_B^2}, \frac{M_N^2}{M_B^2} \right).
\label{pNb}
\eea
\ees
From Eqs.~(\ref{EppnoD}) and (\ref{Np}) we see that in this case $E''_N$ depends only on $\theta_N$, the angle between ${\hat \beta}_B$ and ${\vec {p'_N}}$. The integration differential in Eq.~(\ref{BreffnoD}) thus reduces simply to $d \Omega_{{\hat {p'}}_N} \mapsto 2 \pi d (\cos \theta_N)$.


\begin{thebibliography}{99}



 \bibitem{0NBB}
  G.~Racah,
  ``On the symmetry of particle and antiparticle,''
  Nuovo Cimento  {\bf 14}, 322 (1937)
  doi:10.1007/BF02961321;
  W.~H.~Furry,
  ``On transition probabilities in double beta-disintegration,''
  Phys.\ Rev.\  {\bf 56}, 1184 (1939)
doi:10.1103/PhysRev.56.1184;
H.~Primakoff and S.~P.~Rosen, ``Double beta decay,''
Rep. Prog. Phys. {\bf 22}, 121 (1959);
  ``Nuclear double-beta decay and a new limit on lepton nonconservation,''
  Phys.\ Rev.\  {\bf 184}, 1925 (1969)
   doi:10.1103/PhysRev.184.1925;
  ``Baryon number and lepton number conservation laws,''
  Annu.\ Rev.\ Nucl.\ Part.\ Sci.\  {\bf 31}, 145 (1981)
doi:10.1146/annurev.ns.31.120181.001045;
  J.~Schechter and J.~W.~F.~Valle,
  ``Neutrinoless double beta decay in $SU(2) x U(1)$ theories,''
  Phys.\ Rev.\ D {\bf 25}, 2951 (1982)
  doi:10.1103/PhysRevD.25.2951;
 M.~Doi, T.~Kotani and E.~Takasugi,
  ``Double beta decay and Majorana neutrino,''
  Prog.\ Theor.\ Phys.\ Suppl.\  {\bf 83}, 1 (1985)
   doi:10.1143/PTPS.83.1;
     S.~R.~Elliott and J.~Engel,
  ``Double beta decay,''
  J.\ Phys.\ G  {\bf 30}, R183 (2004)
  doi:10.1088/0954-3899/30/9/R01
  [hep-ph/0405078].
  V.~A.~Rodin, A.~Faessler, F.~\v{S}imkovic and P.~Vogel,
  ``Assessment of uncertainties in QRPA $0\nu\beta\beta$-decay nuclear matrix elements,''
  Nucl.\ Phys.\ A {\bf 766}, 107 (2006)
  Erratum: [Nucl.\ Phys.\ A {\bf 793}, 213 (2007)]
  doi:10.1016/j.nuclphysa.2005.12.004
  [arXiv:0706.4304 [nucl-th]].




\bibitem{RMDs}
 L.~S.~Littenberg and R.~E.~Shrock,
  ``Upper bounds on lepton number violating meson decays,''
  Phys.\ Rev.\ Lett.\  {\bf 68}, 443 (1992)
doi:10.1103/PhysRevLett.68.443;
  ``Implications of improved upper bounds on $|\Delta L| = 2$ processes,''
  Phys.\ Lett.\ B {\bf 491}, 285 (2000)
 doi:10.1016/S0370-2693(00)01041-8
  [hep-ph/0005285];
 C.~Dib, V.~Gribanov, S.~Kovalenko and I.~Schmidt,
  ``K meson neutrinoless double muon decay as a probe of neutrino masses and mixings,''
  Phys.\ Lett.\ B {\bf 493}, 82 (2000)
  doi:10.1016/S0370-2693(00)01134-5
  [hep-ph/0006277];
  A.~Ali, A.~V.~Borisov and N.~B.~Zamorin,
  ``Majorana neutrinos and same sign dilepton production at LHC and in rare meson decays,''
  Eur.\ Phys.\ J.\ C {\bf 21}, 123 (2001)
  doi:10.1007/s100520100702
  [hep-ph/0104123];
  M.~A.~Ivanov and S.~G.~Kovalenko,
  ``Hadronic structure aspects of $K^{+} \to \pi^- + l^+_1 + l^+_2$ decays,''
  Phys.\ Rev.\ D {\bf 71}, 053004 (2005)
 doi:10.1103/PhysRevD.71.053004
  [hep-ph/0412198];
  A.~de Gouvea and J.~Jenkins,
  ``Survey of lepton number violation via effective operators,''
  Phys.\ Rev.\ D {\bf 77}, 013008 (2008)
 doi:10.1103/PhysRevD.77.013008
 [arXiv:0708.1344 [hep-ph]];
  N.~Quintero, G.~L\'opez Castro and D.~Delepine,
  ``Lepton number violation in top quark and neutral B meson decays,''
  Phys.\ Rev.\ D {\bf 84}, 096011 (2011)
  Erratum: [Phys.\ Rev.\ D {\bf 86}, 079905 (2012)]
  doi:10.1103/PhysRevD.86.079905, 10.1103/PhysRevD.84.096011
  [arXiv:1108.6009 [hep-ph]];
  G.~L.~Castro and N.~Quintero,
  ``Bounding resonant Majorana neutrinos from four-body B and D decays,''
  Phys.\ Rev.\ D {\bf 87}, 077901 (2013)
  doi:10.1103/PhysRevD.87.077901
  [arXiv:1302.1504 [hep-ph]];
  A.~Abada, A.~M.~Teixeira, A.~Vicente and C.~Weiland,
  ``Sterile neutrinos in leptonic and semileptonic decays,''
  JHEP {\bf 1402}, 091 (2014)
  doi:10.1007/JHEP02(2014)091
  [arXiv:1311.2830 [hep-ph]];
  Y.~Wang, S.~S.~Bao, Z.~H.~Li, N.~Zhu and Z.~G.~Si,
  ``Study Majorana neutrino contribution to B-meson semi-leptonic rare decays,''
  Phys.\ Lett.\ B {\bf 736}, 428 (2014)
  doi:10.1016/j.physletb.2014.08.006
  [arXiv:1407.2468 [hep-ph]].

  \bibitem{HKS}
  J.~C.~Helo, S.~Kovalenko and I.~Schmidt,
  ``Sterile neutrinos in lepton number and lepton flavor violating decays,''
  Nucl.\ Phys.\ B {\bf 853}, 80 (2011)
  doi:10.1016/j.nuclphysb.2011.07.020
  [arXiv:1005.1607 [hep-ph]].

   \bibitem{Atre}
  A.~Atre, T.~Han, S.~Pascoli and B.~Zhang,
  ``The search for heavy Majorana neutrinos,''
  JHEP {\bf 0905}, 030 (2009)
  doi:10.1088/1126-6708/2009/05/030
  [arXiv:0901.3589 [hep-ph]].

 \bibitem{CDKK}
  G.~Cveti\v{c}, C.~Dib, S.~K.~Kang and C.~S.~Kim,
  ``Probing Majorana neutrinos in rare $K$ and $D, ~D_s, B, B_c$ meson decays,''
  Phys.\ Rev.\ D {\bf 82}, 053010 (2010)
   doi:10.1103/PhysRevD.82.053010
  [arXiv:1005.4282 [hep-ph]].

\bibitem{CDK}
  G.~Cveti\v{c}, C.~Dib and C.~S.~Kim,
  ``Probing Majorana neutrinos in rare $\pi^+ \to e^+ e^+ \mu^- \nu$ decays,''
  JHEP {\bf 1206}, 149 (2012)
  [arXiv:1203.0573 [hep-ph]].

 \bibitem{CKZ}
  G.~Cveti\v{c}, C.~S.~Kim and J.~Zamora-Sa\'a,
  ``CP violations in $\pi^{\pm}$ meson decay,''
  J.\ Phys.\ G {\bf 41}, 075004 (2014)
  doi:10.1088/0954-3899/41/7/075004
  [arXiv:1311.7554 [hep-ph]].

\bibitem{symm}
  G.~Cveti\v{c}, C.~Dib, C.~S.~Kim and J.~Zamora-Sa\'a,
  ``Probing the Majorana neutrinos and their CP violation in decays of charged scalar mesons $\pi, K, D, D_s, B, B_c$,''
Symmetry {\bf 7}, 726 (2015)
 doi:10.3390/sym7020726
  [arXiv:1503.01358 [hep-ph]].

\bibitem{Quint}
  D.~Milanes, N.~Quintero and C.~E.~Vera,
  ``Sensitivity to Majorana neutrinos in $\Delta L=2$ decays of $B_c$ meson at LHCb,''
  Phys.\ Rev.\ D {\bf 93}, no. 9, 094026 (2016)
  doi:10.1103/PhysRevD.93.094026
  [arXiv:1604.03177 [hep-ph]].

\bibitem{Mand}
  S.~Mandal and N.~Sinha,
  ``Favoured $B_c$ decay modes to search for a Majorana neutrino,''
  Phys.\ Rev.\ D {\bf 94}, no. 3, 033001 (2016)
  doi:10.1103/PhysRevD.94.033001
  [arXiv:1602.09112 [hep-ph]].
  
\bibitem{GKS}
  V.~Gribanov, S.~Kovalenko and I.~Schmidt,
  ``Sterile neutrinos in tau lepton decays,''
  Nucl.\ Phys.\ B {\bf 607}, 355 (2001)
  doi:10.1016/S0550-3213(01)00169-9
  [hep-ph/0102155].

\bibitem{tau}
  G.~Cveti\v{c}, C.~Dib, C.~S.~Kim and J.~D.~Kim,
  ``On lepton flavor violation in tau decays,''
  Phys.\ Rev.\ D {\bf 66}, 034008 (2002)
  Erratum: [Phys.\ Rev.\ D {\bf 68}, 059901 (2003)]
  doi:10.1103/PhysRevD.66.034008, 10.1103/PhysRevD.68.059901
  [hep-ph/0202212];
  J.~C.~Helo, S.~Kovalenko and I.~Schmidt,
  ``On sterile neutrino mixing with $\nu_{\tau}$,''
  Phys.\ Rev.\ D {\bf 84}, 053008 (2011)
  doi:10.1103/PhysRevD.84.053008
  [arXiv:1105.3019 [hep-ph]];
  J.~Zamora-Sa\'a,
  ``Resonant $CP$ violation in rare $\tau^{\pm}$ decays,''
  JHEP {\bf 1705}, 110 (2017)
  doi:10.1007/JHEP05(2017)110
  [arXiv:1612.07656 [hep-ph]].
  N.~Shimizu [Belle Collaboration],
  ``New Physics search in rare $\tau$ decays at Belle and prospects at Belle II,''
  PoS FPCP {\bf 2016}, 022 (2017).


\bibitem{scatt1}
  W.~-Y.~Keung and G.~Senjanovi\'c,
  ``Majorana neutrinos and the production of the right-handed charged gauge boson,''
  Phys.\ Rev.\ Lett.\  {\bf 50}, 1427 (1983)
   doi:10.1103/PhysRevLett.50.1427;
  V.~Tello, M.~Nemev\v{s}ek, F.~Nesti, G.~Senjanovi\'c and F.~Vissani,
  ``Left-right symmetry: from LHC to Neutrinoless double beta decay,''
  Phys.\ Rev.\ Lett.\  {\bf 106}, 151801 (2011)
 doi:10.1103/PhysRevLett.106.151801
 [arXiv:1011.3522 [hep-ph]];
  M.~Nemev\v{s}ek, F.~Nesti, G.~Senjanovi\'c and V.~Tello,
  ``Neutrinoless double beta decay: low left-right symmetry scale?,''
  arXiv:1112.3061 [hep-ph];
  S.~Kovalenko, Z.~Lu and I.~Schmidt,
  ``Lepton number violating processes mediated by Majorana neutrinos at hadron colliders,''
  Phys.\ Rev.\ D {\bf 80}, 073014 (2009)
  doi:10.1103/PhysRevD.80.073014
  [arXiv:0907.2533 [hep-ph]];
  G.~Senjanovi\'c,
  ``Neutrino mass: From LHC to grand unification,''
 Riv.\ Nuovo Cim.\  {\bf 34}, 1 (2011)
  doi:10.1393/ncr/i2011-10061-8;
  C.~Y.~Chen and P.~S.~Bhupal Dev,
  ``Multi-lepton collider signatures of heavy Dirac and Majorana neutrinos,''
  Phys.\ Rev.\ D {\bf 85}, 093018 (2012)
  doi:10.1103/PhysRevD.85.093018
  [arXiv:1112.6419 [hep-ph]];
  C.~Y.~Chen, P.~S.~Bhupal Dev and R.~N.~Mohapatra,
  ``Probing Heavy-Light Neutrino Mixing in Left-Right Seesaw Models at the LHC,''
  Phys.\ Rev.\ D {\bf 88}, 033014 (2013)
 doi:10.1103/PhysRevD.88.033014
  [arXiv:1306.2342 [hep-ph]];
  P.~S.~Bhupal Dev, A.~Pilaftsis and U.~k.~Yang,
  ``New Production Mechanism for Heavy Neutrinos at the LHC,''
  Phys.\ Rev.\ Lett.\  {\bf 112}, 081801 (2014)
 doi:10.1103/PhysRevLett.112.081801
  [arXiv:1308.2209 [hep-ph]];
  A.~Das and N.~Okada,
  ``Inverse seesaw neutrino signatures at the LHC and ILC,''
  Phys.\ Rev.\ D {\bf 88}, 113001 (2013)
  doi:10.1103/PhysRevD.88.113001
  [arXiv:1207.3734 [hep-ph]];
  A.~Das, P.~S.~Bhupal Dev and N.~Okada,
  ``Direct bounds on electroweak scale pseudo-Dirac neutrinos from $\sqrt s=8$ TeV LHC data,''
  Phys.\ Lett.\ B {\bf 735}, 364 (2014)
 doi:10.1016/j.physletb.2014.06.058
  [arXiv:1405.0177 [hep-ph]];
  D.~Alva, T.~Han and R.~Ruiz,
  ``Heavy Majorana neutrinos from $W\gamma$ fusion at hadron colliders,''
  JHEP {\bf 1502}, 072 (2015)
  doi:10.1007/JHEP02(2015)072
  [arXiv:1411.7305 [hep-ph]];
  A.~Das and N.~Okada,
  ``Improved bounds on the heavy neutrino productions at the LHC,''
  Phys.\ Rev.\ D {\bf 93}, no. 3, 033003 (2016)
  doi:10.1103/PhysRevD.93.033003
  [arXiv:1510.04790 [hep-ph]];
  ``Bounds on heavy Majorana neutrinos in type-I seesaw and implications for collider searches,''
  arXiv:1702.04668 [hep-ph];
  C.~Degrande, O.~Mattelaer, R.~Ruiz and J.~Turner,
  ``Fully-automated precision predictions for heavy neutrino production mechanisms at hadron colliders,''
  Phys.\ Rev.\ D {\bf 94}, no. 5, 053002 (2016)
  doi:10.1103/PhysRevD.94.053002
  [arXiv:1602.06957 [hep-ph]];
  A.~Das, P.~Konar and S.~Majhi,
  ``Production of Heavy neutrino in next-to-leading order QCD at the LHC and beyond,''
  JHEP {\bf 1606}, 019 (2016)
  doi:10.1007/JHEP06(2016)019
  [arXiv:1604.00608 [hep-ph]];
  A.~Das,
  ``Pair production of heavy neutrinos in next-to-leading order QCD at the hadron colliders in the inverse seesaw framework,''
  arXiv:1701.04946 [hep-ph].



\bibitem{scatt2}
  W.~Buchm\"uller and C.~Greub,
  ``Heavy Majorana neutrinos in electron - positron and electron - proton collisions,''
  Nucl.\ Phys.\ B {\bf 363}, 345 (1991)
  doi:10.1016/0550-3213(91)80024-G;
  M.~Kohda, H.~Sugiyama and K.~Tsumura,
  ``Lepton number violation at the LHC with leptoquark and diquark,''
  Phys.\ Lett.\ B {\bf 718}, 1436 (2013)
 doi:10.1016/j.physletb.2012.12.048
  [arXiv:1210.5622 [hep-ph]].

\bibitem{scatt3}
  J.~C.~Helo, M.~Hirsch and S.~Kovalenko,
  ``Heavy neutrino searches at the LHC with displaced vertices,''
  Phys.\ Rev.\ D {\bf 89}, 073005 (2014)
  Erratum: [Phys.\ Rev.\ D {\bf 93}, no. 9, 099902 (2016)]
  doi:10.1103/PhysRevD.89.073005, 10.1103/PhysRevD.93.099902
  [arXiv:1312.2900 [hep-ph]].



\bibitem{KimLHC}
  C.~O.~Dib and C.~S.~Kim,
  ``Discovering sterile neutrinos lighter than $M_W$ at the LHC,''
  Phys.\ Rev.\ D {\bf 92}, no. 9, 093009 (2015)
  doi:10.1103/PhysRevD.92.093009
  [arXiv:1509.05981 [hep-ph]];
  C.~O.~Dib, C.~S.~Kim, K.~Wang and J.~Zhang,
  ``Distinguishing Dirac/Majorana Sterile Neutrinos at the LHC,''
  Phys.\ Rev.\ D {\bf 94}, no. 1, 013005 (2016)
  doi:10.1103/PhysRevD.94.013005
  [arXiv:1605.01123 [hep-ph]];
  C.~O.~Dib, C.~S.~Kim and K.~Wang,
  ``Signatures of Dirac and Majorana sterile neutrinos in trilepton events at the LHC,''
    Phys.\ Rev.\ D {\bf 95}, no. 11, 115020 (2017)
  doi:10.1103/PhysRevD.95.115020
  [arXiv:1703.01934 [hep-ph]];
  C.~O.~Dib, C.~S.~Kim and K.~Wang,
  ``Search for heavy Sterile neutrinos in trileptons at the LHC,''
  arXiv:1703.01936 [hep-ph];
  A.~Das, P.~S.~B.~Dev and C.~S.~Kim,
  ``Constraining sterile neutrinos from precision Higgs data,''
 Phys.\ Rev.\ D {\bf 95}, no. 11, 115013 (2017)
  doi:10.1103/PhysRevD.95.115013
  [arXiv:1704.00880 [hep-ph]];
  A.~Das, Y.~Gao and T.~Kamon,
  ``Heavy Neutrino Search via the Higgs boson at the LHC,''
  arXiv:1704.00881 [hep-ph].


  \bibitem{Pontecorvo}
  B.~Pontecorvo,
  ``Inverse beta processes and nonconservation of lepton charge,''
  Zh.\ Eksp.\ Teor.\ Fiz.\  {\bf 34}, 247 (1957) [Sov.\ Phys.\ JETP {\bf 7}, 172 (1958)];
  ``Neutrino experiments and the problem of conservation of leptonic charge,''
 Zh.\ Eksp.\ Teor.\ Fiz.\  {\bf 53}, 1717 (1967)  [Sov.\ Phys.\ JETP {\bf 26}, 984 (1968)].

 \bibitem{oscatm}
  Y.~Fukuda {\it et al.}  [Super-Kamiokande Collaboration],
  ``Evidence for oscillation of atmospheric neutrinos,''
  Phys.\ Rev.\ Lett.\  {\bf 81}, 1562 (1998)
  doi:10.1103/PhysRevLett.81.1562
  [hep-ex/9807003].

\bibitem{oscsol}
  Q.~R.~Ahmad {\it et al.}  [SNO Collaboration],
  ``Direct evidence for neutrino flavor transformation from neutral current interactions in the Sudbury Neutrino Observatory,''
  Phys.\ Rev.\ Lett.\  {\bf 89}, 011301 (2002)
   doi:10.1103/PhysRevLett.89.011301
  [nucl-ex/0204008];
  P.~Lipari,
  ``CP violation effects and high-energy neutrinos,''
  Phys.\ Rev.\ D {\bf 64}, 033002 (2001)
 doi:10.1103/PhysRevD.64.033002
  [hep-ph/0102046];
  Z.~Rahman, A.~Dasgupta and R.~Adhikari,
  ``Discovery reach of CP violation in neutrino oscillation experiments with standard and non-standard interactions,''
  arXiv:1210.2603 [hep-ph].
  ``Which baseline for neutrino factory could be better for discovering CP violation in neutrino oscillation for standard and non-standard interactions?,''
  arXiv:1210.4801 [hep-ph].

\bibitem{oscnuc}
  K.~Eguchi {\it et al.}  [KamLAND Collaboration],
  ``First results from KamLAND: Evidence for reactor anti-neutrino disappearance,''
  Phys.\ Rev.\ Lett.\  {\bf 90}, 021802 (2003)
 doi:10.1103/PhysRevLett.90.021802
  [hep-ex/0212021].

 \bibitem{Boya}
  D.~Boyanovsky,
  ``Nearly degenerate heavy sterile neutrinos in cascade decay: mixing and oscillations,''
  Phys.\ Rev.\ D {\bf 90}, 105024 (2014)
 doi:10.1103/PhysRevD.90.105024
  [arXiv:1409.4265 [hep-ph]].

 \bibitem{CKZosc}
  G.~Cveti\v{c}, C.~S.~Kim, R.~K\"ogerler and J.~Zamora-Sa\'a,
  ``Oscillation of heavy sterile neutrino in decay of $B \to \mu e \pi$,''
  Phys.\ Rev.\ D {\bf 92}, 013015 (2015)
  doi:10.1103/PhysRevD.92.013015
  [arXiv:1505.04749 [hep-ph]].


 \bibitem{oscCP}
  N.~Cabibbo,
  ``Time reversal violation in neutrino oscillation,''
  Phys.\ Lett.\ B {\bf 72}, 333 (1978).
  doi:10.1016/0370-2693(78)90132-6

\bibitem{Lepto}
 M.~A.~Luty,
  ``Baryogenesis via leptogenesis,''
  Phys.\ Rev.\ D {\bf 45}, 455 (1992)
  doi:10.1103/PhysRevD.45.455;
  L.~Covi, E.~Roulet and F.~Vissani,
  ``CP violating decays in leptogenesis scenarios,''
  Phys.\ Lett.\ B {\bf 384}, 169 (1996)
  doi:10.1016/0370-2693(96)00817-9
  [hep-ph/9605319].


\bibitem{Pilaftsis}
  A.~Pilaftsis,
  ``CP violation and baryogenesis due to heavy Majorana neutrinos,''
  Phys.\ Rev.\ D {\bf 56}, 5431 (1997)
  doi:10.1103/PhysRevD.56.5431
  [hep-ph/9707235];
  S.~Bray, J.~S.~Lee and A.~Pilaftsis,
  ``Resonant CP violation due to heavy neutrinos at the LHC,''
  Nucl.\ Phys.\ B {\bf 786}, 95 (2007)
   doi:10.1016/j.nuclphysb.2007.07.002
  [hep-ph/0702294 [HEP-PH]].


  \bibitem{CKZ2}
  G.~Cveti\v{c}, C.~S.~Kim and J.~Zamora-Sa\'a,
  ``CP violation in lepton number violating semihadronic decays of $K,D,D_s,B,B_c$,''
  Phys.\ Rev.\ D {\bf 89}, 093012 (2014)
  doi:10.1103/PhysRevD.89.093012
  [arXiv:1403.2555 [hep-ph]].

\bibitem{DCK}
  C.~O.~Dib, M.~Campos and C.~S.~Kim,
  ``CP violation with Majorana neutrinos in $K$ meson decays,''
  JHEP {\bf 1502}, 108 (2015)
 doi:10.1007/JHEP02(2015)108
  [arXiv:1403.8009 [hep-ph]].

\bibitem{nuMSM}
  T.~Asaka, S.~Blanchet and M.~Shaposhnikov,
  ``The $\nu$MSM, dark matter and neutrino masses,''
  Phys.\ Lett.\ B {\bf 631}, 151 (2005)
  doi:10.1016/j.physletb.2005.09.070
  [hep-ph/0503065];
  T.~Asaka and M.~Shaposhnikov,
  ``The $\nu$MSM, dark matter and baryon asymmetry of the universe,''
  Phys.\ Lett.\ B {\bf 620}, 17 (2005)
  doi:10.1016/j.physletb.2005.06.020
  [hep-ph/0505013].


\bibitem{Shapo}
 D.~Gorbunov and M.~Shaposhnikov,
  ``How to find neutral leptons of the $\nu$MSM?,''
  JHEP {\bf 0710}, 015 (2007)
  Erratum: [JHEP {\bf 1311}, 101 (2013)]
  doi:10.1007/JHEP11(2013)101, 10.1088/1126-6708/2007/10/015
  [arXiv:0705.1729 [hep-ph]];
  A.~Boyarsky, O.~Ruchayskiy and M.~Shaposhnikov,
  ``The role of sterile neutrinos in cosmology and astrophysics,''
  Annu.\ Rev.\ Nucl.\ Part.\ Sci.\  {\bf 59}, 191 (2009)
  doi:10.1146/annurev.nucl.010909.083654
  [arXiv:0901.0011 [hep-ph]];
  L.~Canetti, M.~Drewes and M.~Shaposhnikov,
  ``Sterile neutrinos as the origin of dark and baryonic matter,''
  Phys.\ Rev.\ Lett.\  {\bf 110}, 061801 (2013)
 doi:10.1103/PhysRevLett.110.061801
  [arXiv:1204.3902 [hep-ph]];
  L.~Canetti, M.~Drewes, T.~Frossard and M.~Shaposhnikov,
  ``Dark matter, baryogenesis and neutrino oscillations from right handed neutrinos,''
  Phys.\ Rev.\ D {\bf 87}, 093006 (2013)
 doi:10.1103/PhysRevD.87.093006
  [arXiv:1208.4607 [hep-ph]].


\bibitem{lsseesaw}
L.~Canetti, M.~Drewes and B.~Garbrecht,
``Probing leptogenesis with GeV-scale sterile neutrinos at LHCb and Belle II,''
Phys.\ Rev.\ D {\bf 90}, 125005 (2014)
 doi:10.1103/PhysRevD.90.125005
  [arXiv:1404.7114 [hep-ph]];
  M.~Drewes and B.~Garbrecht,
``Experimental and cosmological constraints on heavy neutrinos,''
arXiv:1502.00477 [hep-ph];
  G.~Moreno and J.~Zamora-Sa\'a,
  ``Rare meson decays with three pairs of quasi-degenerate heavy neutrinos,''
  Phys.\ Rev.\ D {\bf 94}, no. 9, 093005 (2016)
  doi:10.1103/PhysRevD.94.093005
  [arXiv:1606.08820 [hep-ph]].


\bibitem{seesaw}
  P.~Minkowski,
  ``$\mu \to e \gamma$ at a rate of one out of $10^9$ muon decays?,''
  Phys.\ Lett.\ B {\bf 67}, 421 (1977)
 doi:10.1016/0370-2693(77)90435-X;
M.~Gell-Mann, P.~Ramond and R.~Slansky, in Sanibel Conference,
``The family group in Grand Unified Theories,'' Febr.~1979, Report No.~CALT-68-709, reprinted in hep-ph/9809459;
``Complex spinors and unified theories,'' Print 80-0576,
published in: D.~Freedman et al. (Eds.), {\it Supergravity}, North-Holland, Amsterdam, 1979;
  T.~Yanagida,
  ``Horizontal symmetry and masses of neutrinos,''
  Conf.\ Proc.\ C {\bf 7902131}, 95 (1979);
S.~L.~Glashow, in: M.~Levy et al. (Eds.), {\it Quarks and Leptons}, Cargese,
Plenum, New York, 1980, p.~707;
  R.~N.~Mohapatra and G.~Senjanovi\'c,
  ``Neutrino mass and spontaneous parity violation,''
  Phys.\ Rev.\ Lett.\  {\bf 44}, 912 (1980).
  doi:10.1103/PhysRevLett.44.912

\bibitem{WWMMD}
  D.~Wyler and L.~Wolfenstein,
  ``Massless neutrinos in left-right symmetric models,''
  Nucl.\ Phys.\ B {\bf 218}, 205 (1983)
 doi:10.1016/0550-3213(83)90482-0;
  E.~Witten,
  ``Symmetry breaking patterns in superstring models,''
  Nucl.\ Phys.\ B {\bf 258}, 75 (1985)
  doi:10.1016/0550-3213(85)90603-0;
  R.~N.~Mohapatra and J.~W.~F.~Valle,
  ``Neutrino mass and baryon number nonconservation in superstring models,''
  Phys.\ Rev.\ D {\bf 34}, 1642 (1986)
  doi:10.1103/PhysRevD.34.1642;
  M.~Malinsky, J.~C.~Romao and J.~W.~F.~Valle,
  ``Novel supersymmetric $SO(10)$ seesaw mechanism,''
  Phys.\ Rev.\ Lett.\  {\bf 95}, 161801 (2005)
 doi:10.1103/PhysRevLett.95.161801
  [hep-ph/0506296];
  P.~S.~Bhupal Dev and R.~N.~Mohapatra,
  ``TeV scale inverse seesaw in $SO(10)$ and leptonic non-unitarity effects,''
  Phys.\ Rev.\ D {\bf 81}, 013001 (2010)
  doi:10.1103/PhysRevD.81.013001
  [arXiv:0910.3924 [hep-ph]];
  P.~S.~Bhupal Dev and A.~Pilaftsis,
  ``Minimal radiative neutrino mass mechanism for inverse seesaw models,''
  Phys.\ Rev.\ D {\bf 86}, 113001 (2012)
  doi:10.1103/PhysRevD.86.113001
  [arXiv:1209.4051 [hep-ph]];
  C.~H.~Lee, P.~S.~Bhupal Dev and R.~N.~Mohapatra,
  ``Natural TeV-scale left-right seesaw mechanism for neutrinos and experimental tests,''
  Phys.\ Rev.\ D {\bf 88}, 093010 (2013)
 doi:10.1103/PhysRevD.88.093010
  [arXiv:1309.0774 [hep-ph]].

\bibitem{HeAAS}
  F.~del Aguila, J.~A.~Aguilar-Saavedra, J.~de Blas and M.~Zralek,
  ``Looking for signals beyond the neutrino Standard Model,''
  Acta Phys.\ Polon.\ B {\bf 38}, 3339 (2007)
  [arXiv:0710.2923 [hep-ph]];
  X.~G.~He, S.~Oh, J.~Tandean and C.~C.~Wen,
  ``Large mixing of light and heavy neutrinos in seesaw models and the LHC,''
  Phys.\ Rev.\ D {\bf 80}, 073012 (2009)
  doi:10.1103/PhysRevD.80.073012
  [arXiv:0907.1607 [hep-ph]].

\bibitem{KS}
  J.~Kersten and A.~Y.~Smirnov,
  ``Right-handed neutrinos at CERN LHC and the mechanism of neutrino mass generation,''
  Phys.\ Rev.\ D {\bf 76}, 073005 (2007)
 doi:10.1103/PhysRevD.76.073005
  [arXiv:0705.3221 [hep-ph]].

\bibitem{AMP}
  A.~Ibarra, E.~Molinaro and S.~T.~Petcov,
  ``TeV scale see-saw mechanisms of neutrino mass generation, the Majorana nature of the heavy singlet neutrinos and $0\nu\beta\beta$-decay,''
  JHEP {\bf 1009}, 108 (2010)
  doi:10.1007/JHEP09(2010)108
  [arXiv:1007.2378 [hep-ph]].

  
\bibitem{NSZ}
  M.~Nemev\v{s}ek, G.~Senjanovi\'c and Y.~Zhang,
  ``Warm dark matter in low scale left-right theory,''
  JCAP {\bf 1207}, 006 (2012)
  doi:10.1088/1475-7516/2012/07/006
  [arXiv:1205.0844 [hep-ph]].

\bibitem{Shrock}
  T.~Appelquist and R.~Shrock,
  ``Neutrino masses in theories with dynamical electroweak symmetry breaking,''
  Phys.\ Lett.\ B {\bf 548}, 204 (2002)
  doi:10.1016/S0370-2693(02)02854-X
  [hep-ph/0204141];
  ``Dynamical symmetry breaking of extended gauge symmetries,''
  Phys.\ Rev.\ Lett.\  {\bf 90}, 201801 (2003)
  doi:10.1103/PhysRevLett.90.201801
  [hep-ph/0301108];
  ``Fermion masses and mixing in extended technicolor models,''
  Phys.\ Rev.\ D {\bf 69}, 015002 (2004)
  doi:10.1103/PhysRevD.69.015002
  [hep-ph/0308061].
  

\bibitem{BdecBII}
  G.~Cveti\v{c} and C.~S.~Kim,
  ``Rare decays of B mesons via on-shell sterile neutrinos,''
  Phys.\ Rev.\ D {\bf 94}, no. 5, 053001 (2016)
  Erratum: [Phys.\ Rev.\ D {\bf 95}, no. 3, 039901 (2017)]
  doi:10.1103/PhysRevD.95.039901, 10.1103/PhysRevD.94.053001
  [arXiv:1606.04140 [hep-ph]].


\bibitem{Sheldon}
  Sheldon L.~Stone, private communication.

\bibitem{Belle-II}
  T.~Aushev {\it et al.},
  ``Physics at Super B Factory,''
  arXiv:1002.5012 [hep-ex];
 T.~Abe {\it et al.} [Belle-II Collaboration],
  ``Belle II technical design report,''
  arXiv:1011.0352 [physics.ins-det].



\bibitem{BelleUB}
  D.~Liventsev {\it et al.} [Belle Collaboration],
  ``Search for heavy neutrinos at Belle,''
  Phys.\ Rev.\ D {\bf 87}, no. 7, 071102 (2013)
  doi:10.1103/PhysRevD.87.071102
  [arXiv:1301.1105 [hep-ex], v3 with Erratum (2017)].

\bibitem{LHCba1}
  R.~Aaij {\it et al.} [LHCb Collaboration],
  ``Searches for Majorana neutrinos in $B^-$ decays,''
  Phys.\ Rev.\ D {\bf 85}, 112004 (2012)
  doi:10.1103/PhysRevD.85.112004
  [arXiv:1201.5600 [hep-ex]];
  ``Search for Majorana neutrinos in $B^- \to \pi^+\mu^-\mu^-$ decays,''
  Phys.\ Rev.\ Lett.\  {\bf 112}, no. 13, 131802 (2014)
  doi:10.1103/PhysRevLett.112.131802
  [arXiv:1401.5361 [hep-ex]].

\bibitem{LHCba2}
  B.~Shuve and M.~E.~Peskin,
  ``Revision of the LHCb limit on Majorana neutrinos,''
  Phys.\ Rev.\ D {\bf 94}, no. 11, 113007 (2016)
  doi:10.1103/PhysRevD.94.113007
  [arXiv:1607.04258 [hep-ph]].



\bibitem{PDG2016}
  C.~Patrignani {\it et al.} [Particle Data Group Collaboration],
  ``Review of particle physics,''
  Chin.\ Phys.\ C {\bf 40}, no. 10, 100001 (2016).
  doi:10.1088/1674-1137/40/10/100001


\bibitem{Kangetal}
  X.~W.~Kang, B.~Kubis, C.~Hanhart and U.~G.~Meißner,
  ``$B_{l4}$ decays and the extraction of $|V_{ub}|$,''
  Phys.\ Rev.\ D {\bf 89}, 053015 (2014)
  doi:10.1103/PhysRevD.89.053015
  [arXiv:1312.1193 [hep-ph]].

\bibitem{CLN}
  I.~Caprini, L.~Lellouch and M.~Neubert,
  ``Dispersive bounds on the shape of ${\overline B} \to D^{(*)} \ell {\overline \nu}$ form factors,''
  Nucl.\ Phys.\ B {\bf 530}, 153 (1998)
  doi:10.1016/S0550-3213(98)00350-2
  [hep-ph/9712417].

\bibitem{Belle1}
  R.~Glattauer {\it et al.} [Belle Collaboration],
  ``Measurement of the decay $B\to D\ell\nu_\ell$ in fully reconstructed events and determination of the Cabibbo-Kobayashi-Maskawa matrix element $|V_{cb}|$,''
  Phys.\ Rev.\ D {\bf 93}, no. 3, 032006 (2016)
  doi:10.1103/PhysRevD.93.032006
  [arXiv:1510.03657 [hep-ex]].

  \bibitem{CaNeu}
  I.~Caprini and M.~Neubert,
  ``Improved bounds for the slope and curvature of ${\bar B} \to D^{(*)} \ell {\overline \nu}$ form-factors,''
  Phys.\ Lett.\ B {\bf 380}, 376 (1996)
  doi:10.1016/0370-2693(96)00509-6
  [hep-ph/9603414].

\bibitem{NeuPRps}
  M.~Neubert,
  ``Heavy quark symmetry,''
  Phys.\ Rept.\  {\bf 245}, 259 (1994)
  doi:10.1016/0370-1573(94)90091-4
  [hep-ph/9306320].

\bibitem{GiSi}
  F.~J.~Gilman and R.~L.~Singleton,
  ``Analysis of semileptonic decays of mesons containing heavy quarks,''
  Phys.\ Rev.\ D {\bf 41}, 142 (1990).
  doi:10.1103/PhysRevD.41.142

 \bibitem{Belle2}
  W.~Dungel {\it et al.} [Belle Collaboration],
  ``Measurement of the form factors of the decay $B^0 \to D^{*-} \ell^{+} \nu$ and determination of the CKM matrix element $|V_{cb}|$,''
  Phys.\ Rev.\ D {\bf 82}, 112007 (2010)
  doi:10.1103/PhysRevD.82.112007
  [arXiv:1010.5620 [hep-ex]].

\bibitem{CERN-SPS}
  W.~Bonivento {\it et al.},
  ``Proposal to search for heavy neutral leptons at the SPS,''
  arXiv:1310.1762 [hep-ex].


\bibitem{commKim}
  C.~Dib and C.~S.~Kim,
  ``Remarks on the lifetime of sterile neutrinos and the effect on detection of rare meson decays $M^+ \to M^{\prime}-\ell^+\ell^+$,''
  Phys.\ Rev.\ D {\bf 89}, no. 7, 077301 (2014)
  doi:10.1103/PhysRevD.89.077301
  [arXiv:1403.1985 [hep-ph]].

\bibitem{Gronau}
  M.~Gronau, C.~N.~Leung and J.~L.~Rosner,
  ``Extending limits on neutral heavy leptons,''
  Phys.\ Rev.\ D {\bf 29}, 2539 (1984).
  doi:10.1103/PhysRevD.29.2539

\bibitem{LHCC98004}
  R.~Aaij {\it et al.} [LHCb Collaboration],
  ``LHCb : Technical Proposal,'' CERN-LHCC-98-004, 3 March 1998, https://cds.cern.ch/record/622031?ln=en and http://lhcb-tp.web.cern.ch/lhcb-tp/html/lhccframes.htm, Sec.~3.1 there (Overview of the experiment, Particle identification).

\bibitem{VELO}
  R.~Aaij {\it et al.} [LHCb Collaboration],
  ``LHCb detector performance,''
  Int.\ J.\ Mod.\ Phys.\ A {\bf 30}, no. 07, 1530022 (2015)
  doi:10.1142/S0217751X15300227
  [arXiv:1412.6352 [hep-ex]];
  ``LHCb VELO Upgrade Technical Design Report,'' CERN-LHCC-2013-021,  29 Nov.~2013, https://cds.cern.ch/record/1624070, cf.~Fig.~4 there.

\bibitem{RICHdesign}
  R.~Aaij {\it et al.} [LHCb Collaboration],
  ``LHCb PID Upgrade Technical Design Report,'' CERN-LHCC-2013-022, 28 Nov.~2013, https://cds.cern.ch/record/1624074?ln=en, cf.~Fig.~2,20 there.

\bibitem{FC}
  G.~J.~Feldman and R.~D.~Cousins,
  ``A unified approach to the classical statistical analysis of small signals,''
  Phys.\ Rev.\ D {\bf 57}, 3873 (1998)
  doi:10.1103/PhysRevD.57.3873
  [physics/9711021 [physics.data-an]].

\bibitem{AsIsh}
  T.~Asaka and H.~Ishida,
  ``Lepton number violation by heavy Majorana neutrino in $B$ decays,''
  Phys.\ Lett.\ B {\bf 763}, 393 (2016)
  doi:10.1016/j.physletb.2016.10.070
  [arXiv:1609.06113 [hep-ph]].


\bibitem{DELPHI}
  P.~Abreu {\it et al.} [DELPHI Collaboration],
  ``Search for neutral heavy leptons produced in Z decays,''
  Z.\ Phys.\ C {\bf 74}, 57 (1997)
  Erratum: [Z.\ Phys.\ C {\bf 75}, 580 (1997)].
  doi:10.1007/s002880050370

\bibitem{BEBC}
  A.~M.~Cooper-Sarkar {\it et al.} [WA66 Collaboration],
  ``Search for heavy neutrino decays in the BEBC beam dump experiment,''
  Phys.\ Lett.\  {\bf 160B}, 207 (1985).
  doi:10.1016/0370-2693(85)91493-5

\bibitem{NuTeV}
  .~Vaitaitis {\it et al.} [NuTeV and E815 Collaborations],
  ``Search for neutral heavy leptons in a high-energy neutrino beam,''
  Phys.\ Rev.\ Lett.\  {\bf 83}, 4943 (1999)
  doi:10.1103/PhysRevLett.83.4943
  [hep-ex/9908011].

\bibitem{NA3}
  J.~Badier {\it et al.} [NA3 Collaboration],
  ``Mass and lifetime limits on new longlived particles in 300 GeV  $\pi^-$ interactions,''
  Z.\ Phys.\ C {\bf 31}, 21 (1986).
  doi:10.1007/BF01559588

\bibitem{CHARMII}
  P.~Vilain {\it et al.} [CHARM II Collaboration],
  ``Search for heavy isosinglet neutrinos,''
  Phys.\ Lett.\ B {\bf 343}, 453-458 (1995)
  [Phys.\ Lett.\ B {\bf 351}, 387 (1995)].
  doi:10.1016/0370-2693(94)00440-I, 10.1016/0370-2693(94)01422-9


\bibitem{Kova}
  P.~Bene\v{s}, A.~Faessler, F.~\v{S}imkovic and S.~Kovalenko,
  ``Sterile neutrinos in neutrinoless double beta decay,''
  Phys.\ Rev.\ D {\bf 71}, 077901 (2005)
  doi:10.1103/PhysRevD.71.077901
  [hep-ph/0501295].
  A.~Faessler, M.~Gonz\'alez, S.~Kovalenko and F.~\v{S}imkovic,
  ``Arbitrary mass Majorana neutrinos in neutrinoless double beta decay,''
  Phys.\ Rev.\ D {\bf 90}, no. 9, 096010 (2014)
  doi:10.1103/PhysRevD.90.096010
  [arXiv:1408.6077 [hep-ph]].




\end{thebibliography}
\end{document}